\documentclass[twocolumn,showpacs,preprintnumbers,prd,aps,10pt,nofootinbib]{
revtex4-1}
\usepackage{bm}
\usepackage{amssymb,amsmath}
\usepackage{graphicx}
\usepackage{natbib}
\usepackage{program}
\usepackage{color}

\usepackage[breaklinks,colorlinks,citecolor=blue]{hyperref}

\begin{document}
\title[SP--VP]{On the Schr\"odinger-Poisson--Vlasov-Poisson correspondence}

\author{Philip Mocz}
\thanks{Einstein Fellow}
\email{pmocz@astro.princeton.edu}
\affiliation{Department of Astrophysical Sciences, Princeton University, 4 Ivy Lane, Princeton, NJ, 08544, USA}
\author{Lachlan Lancaster}
\affiliation{Department of Astrophysical Sciences, Princeton University, 4 Ivy Lane, Princeton, NJ, 08544, USA}
\author{Anastasia Fialkov}
\affiliation{Harvard-Smithsonian Center for Astrophysics, 60 Garden Street, Cambridge, MA 02138, USA}
\author{Fernando Becerra}
\affiliation{Harvard-Smithsonian Center for Astrophysics, 60 Garden Street, Cambridge, MA 02138, USA}
\author{Pierre-Henri Chavanis}
\affiliation{Laboratoire de Physique Th\'eorique, Universit\'e Paul Sabatier, 118 route de Narbonne 31062 Toulouse, France}

\date{Mar 2018}




\begin{abstract}
The Schr\"odinger-Poisson equations describe the behavior of a superfluid Bose-Einstein condensate under self-gravity with a 3D wave function. As $\hbar/m\to 0$, $m$
being the boson mass, the equations have been postulated to
approximate the collisionless Vlasov-Poisson equations also known as the
collisionless Boltzmann-Poisson equations. The latter describe
collisionless matter with a 6D classical distribution function.  We investigate
the nature of this correspondence with a suite of numerical test problems in 1D,
2D, and 3D along with analytic treatments when possible.   We
demonstrate that, while the density field of the superfluid always shows order
unity oscillations as $\hbar/m\to 0$ due to interference and the uncertainty
principle, the potential field converges to the classical
answer as $(\hbar/m)^{2}$. Thus, any dynamics coupled to the superfluid
potential is expected to recover the classical collisionless limit as
$\hbar/m\to 0$. The quantum superfluid is able to capture rich phenomena such as
multiple phase-sheets, shell-crossings, and warm distributions. Additionally,
the quantum pressure tensor acts as a regularizer of caustics and singularities
in classical solutions. This suggests the exciting prospect of using the
Schr\"odinger-Poisson equations as a low-memory method for approximating the
high-dimensional evolution of the Vlasov-Poisson equations. As a particular
example we consider dark matter composed of ultra-light axions, which in the
classical limit ($\hbar/m\to 0$) is expected to manifest itself as collisionless
cold dark matter.
\end{abstract}
\smallskip
\maketitle

\section{Introduction}\label{sec:intro}

The Schr\"odinger-Poisson (SP) equations describe a wide variety of physical phenomena.
These include optical systems \citep{2016NatCo...713492R}
and semiconductors \citep{harrison2016quantum}.
In astrophysics, the equations have been 
suggested to model a number of theoretical ideas.
One such is hypothesized boson stars \citep{1968PhRv..172.1331K,1969PhRv..187.1767R,2003CQGra..20R.301S}, which could be a source of `exotic' Laser Interferometer Gravitational-Wave Observatory (LIGO) detections 
\citep{2017PhRvD..96b4002S} aside from the standard signatures of expected gravitational wave merger signals of black hole and neutron star binary systems.
A second theoretical model is axion-like dark matter, which, in the limit of an ultralight mass, is known as fuzzy dark matter \citep[FDM,][]{1989PhRvA..39.4207M,1994PhRvD..50.3650S,1994PhRvD..50.3655J,1996PhRvD..53.2236L,2000PhRvL..85.1158H,2000ApJ...534L.127P,2000NewA....5..103G,2002esas.book..165M,2007JCAP...06..025B,2011PhRvD..84d3531C,2014PhRvL.113z1302S,2014NatPh..10..496S,2016PhRvD..94d3513S,2017MNRAS.471.4559M,2017PhRvD..95d3541H,2017arXiv170405057L}. Axion dark matter has been of
considerable interest of late as an alternative to the standard cold dark matter
(CDM) paradigm. 
In FDM, the ultralight scalar bosons form a Bose-Einstein condensate (all particles share the same wave function and quantum mechanical effects become macroscopic).
FDM is expected to match CDM on cosmological scales larger than
few kilo-parsecs (kpc) but may solve the small-scale cosmological problems
associated with scale-free CDM, such as the `cusp-core' problem
\citep{moore1994,flores1994,2010AdAst2010E...5D} 
or `too-big-to-fail' problem \citep{boylan-kolchin2011,2012MNRAS.422.1203B} through
macroscopic quantum effects. For example, axions with mass $m\sim 10^{-22}$ eV
transform dark matter halo cusps into kpc-scale soliton cores. Axion dark matter
particles are constrained to have a mass $m\gtrsim 10^{-24}$~eV in order to not
erase the observed large-scale cosmic structure \citep{2015PhRvD..91j3512H}. At
axion masses of $m\sim 10^{-22}$~eV galactic halos will exhibit kpc scale
quantum structures due to the de Broglie wavelength \citep{2011PhRvD..84d3531C}.
At larger axion masses $m\gtrsim 10^{-20}$~eV, the behavior is assumed to start
recovering classical CDM on most cosmological scales of interest, and therefore
it would be difficult to constrain the axion mass above this value based on
structure formation today. These higher mass axions are also well motivated from
particle physics such as the QCD axion
\citep{1977PhRvL..38.1440P} which has been proposed as a theoretical solution to
the strong-CP problem and would have a mass of around $m\sim
10^{-6}$--$10^{-3}$~eV (although lower masses are possible).

The Schr\"odinger equations in these models typically describe bosons sharing a single wave function $\psi(\mathbf{x},t)$,
 and the Poisson equations seed the self-potential (electric or gravitational) that affects the time evolution.
In the equations, the parameter $\hbar/m$, where $m$ is the boson mass, sets the de Broglie wavelength and describes the relative importance of quantum mechanical behavior compared to the classical potential.
Understanding the behavior of the SP equations in the limit as $\hbar/m\to 0$ asks fundamental and practical questions.
As $\hbar\to 0$ one is interested in whether the system leads to the emergence of classical phenomena, in the spirit of the Bohmian interpretation of quantum mechanics.
Alternatively, as $m\to\infty$, one recovers the same limit, so it is a useful way to constrain the boson mass for hypothesized systems such as axion dark matter and to understand limiting behavior.

As $\hbar/m\to 0$, the SP system is expected to approximate the
classical Vlasov-Poisson (VP) equations in some sense, the full theoretical nature of which is not fully understood. 
The VP equations describe the evolution of collisionless classical particles 
governed by the Poisson equation (e.g. self-gravity or electrostatics).
The standard CDM model for example is governed by the VP equations: the dark matter is cold in this case, meaning that initially the particles have a single velocity at any physical location. The VP equations describe the evolution of the 6D phase-space $f(\mathbf{x},\mathbf{v},t)$, thus, for the correspondence to be valid the 3D wave function $\psi(\mathbf{x},t)$ must encode this higher-dimensional information. 
It is not clearly evident that the SP equations should recover the classical VP
limit. 
Classically, the VP equations may have complicated folded phase-sheets in 6D, or
warm distributions, while the SP equations are equivalent,
through the Madelung \cite{1927ZPhy...40..322M} transformation, to a fluid
description with only a single velocity at a given location.

\begin{figure}
\begin{center}
\includegraphics[width=0.5\textwidth]{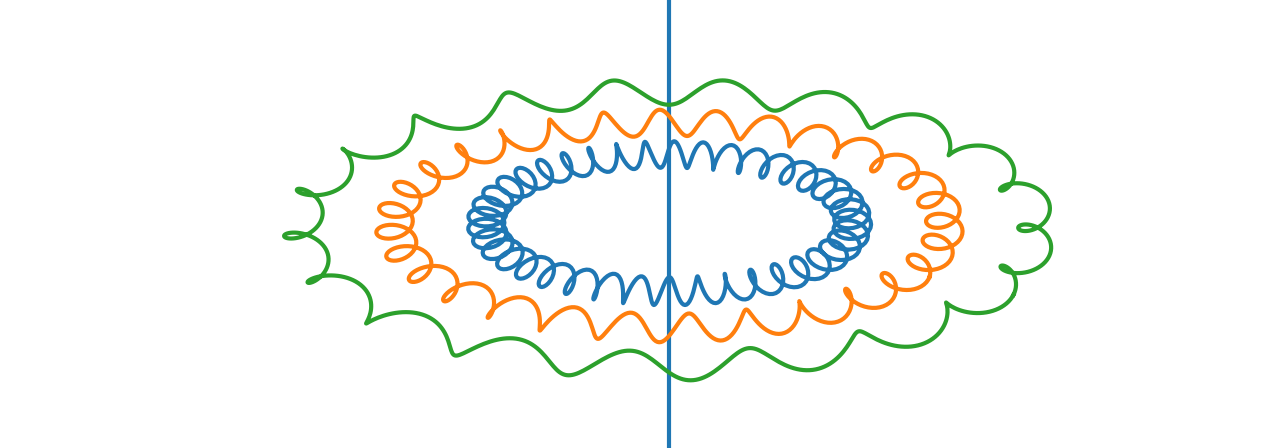}
\end{center}
\caption{A visual aid illustrating how one gets vorticity out of a quantum wave function described by the Schr\"odinger equation.  The vorticity comes from the integral around a loop of the velocity of the fluid. Normally this would be zero as the velocity in a superfluid comes from the gradient of the phase, and thus is curl-free.  However, if the phase is discontinuous at a point or set of points with the density (indicated here by the vertical blue line), then the periodicity of the phase gives rise to quantized vorticity.  The higher the `winding number' or the number of times that the phase goes through a full $2\pi$ radians, the higher the vorticity.  In the figure, each curve is meant to show the evolution of the internal phase of an example field with different winding numbers as you travel around a vortex line. Here we have the winding number set to 20, 30, and 40 for the green, orange, and blue curves respectively.}\label{fig:winding}
\end{figure}

While a complete theoretical understanding of the SP--VP correspondence is lacking, there has been important work demonstrating some of its aspects in 1D and 2D.
The work by \cite{1993ApJ...416L..71W} suggested using the SP equations with artificially small $\hbar$ as a way to simulate the VP equations for CDM, and performed 1D simulations of the Jeans instability which starts from cold initial conditions and exhibits caustics and shell-crossing in its evolution. 
The authors of \cite{1993ApJ...416L..71W} were able to recover a 2D distribution function from the 1D wave function by constructing the Husimi distribution function from $\psi$ (a smoothed version of the Wigner quasiprobability distribution), which was found to resemble the time-evolution of the classical distribution function.
That is, the quantum mechanical results, when smoothed over the local de Broglie wavelength, resemble the classical solution, which is scale-free. We will review this argument in section \ref{sec:bkgd}.
Recently, the authors of \cite{2017arXiv171100140K} have explored the correspondence numerically in 2D with cold cosmological initial conditions for dark matter. 
The work \cite{2017arXiv171100140K} compared cumulants of the distribution function (density, velocity and velocity dispersion) obtained in the SP case through quasi-local manipulations of the wave function, and found excellent qualitative and quantitative agreement (see also the recent work on recovering cumulants in 1D in \cite{2017arXiv171004846G}).
Conceptually this is an interesting and non-trivial result because the VP equations can exhibit vorticity while the SP equations are vortex-free except for degenerate sites of quantized vorticity (``vortex cores'') which we illustrate in Figure \ref{fig:winding}; hence the SP equations recover an effective vorticity through the winding number around the sites where the wave function vanishes.

Without non-local manipulation of the quantum wave function (e.g., through Husimi smoothing)
the density field $\rho=\lvert \psi \rvert^2$ obtained from the
wave function in the Schr\"odinger equation does not recover the classical
density field $\rho=\int f\,d^3v$ from the Vlasov equations.
This is easy to see by considering a distribution which is the superposition of
two Gaussian waves traveling with opposite velocities.
In the classical case, the superposition is still just a Gaussian. 
But in the quantum case, the superposition exhibits interference patterns
of characteristic size the de Broglie wavelength. As $\hbar/m \to 0$, 
the period of the interference oscillations decreases as $\hbar/m$ but the envelope of the wave function remains constant. Therefore the density field always exhibits order unity differences from the classical solution. 
The potential (obtained from the density via the Poisson equation) and the force field (gradient of the potential)
will also show oscillations on the scale of the de Broglie wavelength.
However, fortunately the amplitude of the oscillations in the potential 
is not order unity, rather it is suppressed by a factor of $(\hbar/m)^{2}$ by the $\nabla^2$ operator in the Poisson equation, 
and likewise in the force field the amplitude is suppressed by a factor of $(\hbar/m)$. 
Therefore, these quantities are hypothesized here to converge to the classical solution without non-local manipulation/smoothing, which is a necessary requirement for the SP--VP correspondence to hold under time evolution.
It is further illustrative to consider the time evolution of this simple example of two Gaussians, assuming no self-gravity. In the classical case the two Gaussians will pass through each other and continue traveling with their initial velocity, while in the quantum case the two Gaussians 
will exhibit interference as they pass, but also experience dispersion under
time evolution. As $\hbar/m \to 0$, the dispersion of the Gaussian 
wave packets also goes to $0$, which is necessary for
the SP--VP correspondence to hold.
This example illustrates some of the nature of the SP--VP correspondence as well as why it may be intuitively expected (non-convergence of density field, $(\hbar/m)^{2}$ convergence of potential).

In the present work, we are interested in exploring numerically in 1D, 2D, and 3D the nature of the SP--VP correspondence for complicated test problems which have caustics, shell-crossings, or non-cold initial conditions.
Of great interest is to test whether we can recover the classical potential $V$ in a formal converged sense, with convergence rate faster than $(\hbar/m)^{1}$ (so that the force field is also guaranteed to converge to the classical limit as $\hbar/m\to 0$).
A related question is what happens to nonlinear quantum
structures as $\hbar/m\to 0$.

We would like to understand the SP--VP correspondence under time evolution for several reasons.
First, the correspondence offers a way of understanding the emergence of classical behavior from a quantum system as $\hbar\to 0$.
Additionally, convergence guarantees that one can accurately solve the SP equations as a low-memory method to numerically simulate the rich phase-space structure of the 6D VP equations.
Thirdly, it is a way to learn about the limiting behavior of superfluid Bose-Einstein condensate systems,
for example, in the case of axion dark matter, where the boson mass $m$ is
unknown. FDM--CDM correspondence can be thought of as a special case of the
SP--VP  correspondence discussed here, since taking $\hbar/m\to 0$ is equivalent
to taking the de Broglie wavelength of the superfluid (i.e., the scale at which
the quantum effects of the superfluid are evident) to zero. For the claim that
the classical limit is recovered to be true, it remains to be demonstrated that
the non-smoothed potential 
and force field approach the classical limit in a formal
converged sense so that baryonic matter, which is coupled to the dark matter
only through the gravitational potential, would experience identical forces. 

The paper is organized as follows.
In section~\ref{sec:bkgd} we lay out the theoretical background
for the SP and VP equations.
In section~\ref{sec:method} we describe our numerical simulation method.
We carry out and discuss a number of simulations, including full 3D cosmological simulations of FDM at different boson masses in section~\ref{sec:sims}. These are compared to classical $N$-body simulations of collisionless CDM. 
Our concluding remarks are offered in section~\ref{sec:conc}.
We provide a heuristic discussion in Appendix
\ref{sec_lb} on 
the process of violent relaxation of collisionless self-gravitating systems, 
which have concepts relevant to understand what is happening in our cosmological
simulations of halos.

\section{Theoretical Background}
\label{sec:bkgd}

\subsection{Schr\"odinger-Poisson}

The SP equations describe a self-gravitating quantum superfluid (such as FDM):
\begin{equation}
i\hbar \frac{\partial \psi}{\partial t} =
-\frac{\hbar^2}{2m}\nabla^2\psi + m V\psi,
\end{equation}
\begin{equation}
\nabla^2 V = 4\pi G(\rho-\overline{\rho}),
\end{equation}
where $\psi$ is the wave function describing the scalar field
boson in the nonrelativistic limit, 
$\rho\equiv\lvert\psi\rvert^2$ is the density, $\overline{\rho}$ is the
volume-averaged density, $V$ is the gravitational potential and $m$ is the boson
mass.  It is prudent to note that we are making use of 
the so called `Jeans Swindle' (both here and in Eq.
(\ref{eqn:coVP2})) by sourcing the potential only by the overdensity.

An equivalent formulation of the SP equations is the Madelung \cite{1927ZPhy...40..322M} fluid form, 
which can be useful when interpreting some of the results.
Decomposing the wave function as
\begin{equation}
\psi = \sqrt{\rho} \, {\rm e}^{iS/\hbar}
\label{eqn:decomp}
\end{equation}
and defining a velocity as the gradient of the phase:
\begin{equation}
\mathbf{u} \equiv \frac{\nabla S}{m},
\label{eqn:vmadel}
\end{equation}
the Schr\"odinger equation can then be written as
\begin{equation}
\frac{\partial \rho}{ \partial t}
+ \nabla \cdot \left(
\rho  \mathbf{u}
\right) = 0,
\end{equation}
\begin{equation}
\frac{\partial \mathbf{u}}{\partial t}
+ (\mathbf{u} \cdot\nabla) \mathbf{u}
= -\nabla V - \nabla V_{\rm Q},
\label{eqn:madQ}
\end{equation}
where
\begin{equation}
V_{\rm Q}\equiv
-\frac{\hbar^2}{2m^2}\frac{\nabla^2\sqrt{\rho}}{\sqrt{\rho}}.
\end{equation}
Aside from the quantum potential term $V_{\rm Q}$, the evolution equations
look like that of classical evolution of individual particles under self-gravity, in the spirit of a Bohmian interpretation of quantum mechanics.
The quantum potential can also be written as a quantum pressure tensor
$\mathbb{P}_{\rm Q}$:
\begin{equation}
\nabla V_{\rm Q} = \frac{1}{\rho} \nabla \cdot \mathbb{P}_{\rm
Q},
\end{equation}
\begin{equation}
\mathbb{P}_{\rm Q} \equiv -\frac{\hbar^2}{4m^2}
\rho\nabla\otimes\nabla \ln \rho.
\label{Eq:PQ}
\end{equation}

For $\hbar\rightarrow 0$, we cannot make $V_{\rm Q}=0$ in Eq.~(\ref{eqn:madQ}) otherwise we get the pressureless Euler-Poisson (EP) equations
which develop caustics, i.e., density singularities associated with particle
crossing. As a
result, the EP equations are not defined for all times (they are only valid
until
shell-crossing) contrary to the SP
equations. Therefore, the limit $\hbar\rightarrow 0$ is different from
$\hbar=0$.

\subsection{Vlasov-Poisson}

A classical collisionless fluid (such as CDM), is governed by
the VP equations:
\begin{equation}
\frac{\partial f}{\partial t}
+ \mathbf{v}\cdot \frac{\partial f}{\partial \mathbf{x}}
- \nabla V \cdot \frac{\partial f}{\partial \mathbf{v}} = 0,
\label{eqn:coVP1}
\end{equation}
\begin{equation}
\nabla^2 V = 4\pi G(\rho-\overline{\rho}),
\label{eqn:coVP2}
\end{equation}
where $f=f(\mathbf{x},\mathbf{v},t)$ is the 6D distribution function, 
and the density is given by $\rho = \int f \,d^3v$. We note that the
Vlasov equation is also known alternatively as the
collisionless Boltzmann equation.
The equation is a statement of conservation of phase-space density ($df=0$) and has symplectic structure.

\subsection{Comparison of SP and VP}


While a rigorous proof is lacking \citep{2017arXiv171100140K}, there 
exist good mathematical arguments for why the SP equations may be expected to
recover the classical VP limit as $\hbar/m\to 0$.  On small scales, the SP
system is expected to show quantum phenomenon, such as soliton cores (stable,
ground-state eigenmodes where the uncertainty principle
prevents gravitational collapse; \cite{2014NatPh..10..496S}),
vortex lines and reconnection \citep{2017MNRAS.471.4559M}, interference
patterns, 
quantum tunneling, and non-classical phenomena if there exists jumps in the wave
function phase (e.g. colliding cores can bounce off each other;
\cite{2016PhRvD..94d3513S}). However, on large scales, the quantum behavior is
expected to `average out' to zero and become negligible.
Therefore FDM has a quantum behavior at small scales and
behaves as CDM at large scales.

We note that the SP and VP equations have a scaling symmetry:
\begin{equation}
\{ x,t,\rho,m \} \to \{ \alpha x, 
\beta t, \beta^{-2} \rho, \alpha^{-2}\beta m\}.
\label{eqn:scaling}
\end{equation}
While the VP equations are scale-free,
the SP equations have a single scale (the de Broglie wavelength) set by the value of $\hbar/m$. Thus the SP equations will also naturally become scale free as $\hbar/m \to 0$. 

Below we describe other aspects of the classical and quantum correspondence.

\subsubsection{Jeans equations \& Madelung formalism}

From the  Vlasov equation, we can derive a system of
hydrodynamic equations called the Jeans equations \citep{2008gady.book.....B}. By integrating the Vlasov
equation over velocity, we get the continuity equation (expressing the local
mass conservation):
\begin{equation}
\label{jeans1}
\frac{\partial \rho}{ \partial t}
+ \nabla \cdot \left(
\rho  \langle \mathbf{v} \rangle
\right) = 0,
\end{equation}
where $\langle\cdot\rangle$ is the average over momentum-space.
By multiplying the Vlasov equation by ${\bf v}$ and integrating over
velocity, we obtain the momentum equation:
\begin{equation}
\label{jeans2}
\frac{\partial \langle \mathbf{v} \rangle}{\partial t}
+ \langle \mathbf{v} \rangle \cdot\nabla \langle \mathbf{v} \rangle
= -\nabla V - \frac{1}{\rho} \nabla \cdot\mathbb{P}_{\rm J},
\end{equation}
where 
\begin{equation}
\mathbb{P}_{\rm J} \equiv \rho \sigma_{ij}^2 = \rho \left(\langle v_i v_j\rangle -  \langle v_i \rangle  \langle v_j\rangle \right)
\end{equation} is the stress-tensor. These equations are essentially
those for a compressible fluid which is supported by pressure in the form of a
velocity dispersion. These equations
are not closed. Actually, we can build up an infinite hierarchy of equations by
introducing higher and higher moments of the velocity. In general, there is no
simple way to close this hierarchy of equations except in a particular case.

The VP equations admit a particular solution of the
form
\begin{equation}
\label{jeans3}
f({\bf x},{\bf v},t)=\rho({\bf x},t)\delta({\bf v}-{\bf u}({\bf x},t)).
\end{equation}
This is called the single-speed solution because there is a single velocity
attached to any given $({\bf x},t)$. It is also a zero-pressure solution since
there is no thermal motion. The density $\rho$ and the velocity ${\bf
u}$ satisfy the pressureless EP equations
\begin{equation}
\label{jeans4}
\frac{\partial\rho}{\partial t}+\nabla\cdot (\rho {\bf u})=0,
\end{equation}
\begin{equation}
\label{jeans5}
\frac{\partial {\bf u}}{\partial t}+({\bf u}\cdot \nabla){\bf
u}=-\nabla V,
\end{equation}
\begin{equation}
\nabla^2 V = 4\pi G(\rho-\overline{\rho}).
\label{jeans6}
\end{equation}
These equations are exact. They can be obtained from the Jeans equations
(\ref{jeans1}) and (\ref{jeans2}) by closing the hierarchy with the condition
$\mathbb{P}_{\rm J}=0$ resulting from Eq. (\ref{jeans3}). This corresponds to
the ``dust
model''.  However, there is a well-known difficulty with the solution
(\ref{jeans3}).
Generically, after a finite time the solution of the VP equations becomes
multi-stream because of
particle crossing. This leads to the
formation of caustics (singularities) in the density field at shell-crossing.
This phenomenon renders the pressureless hydrodynamical
description (\ref{jeans4})-(\ref{jeans6}) useless beyond the first time of
crossing when the fast particles cross the slow ones.
Therefore,
after shell-crossing we must come back to the original VP equations or to the
Jeans
equations because we need to account for a velocity dispersion (the velocity
field is multi-valued). We can heuristically cure the problems of the EP
equations by introducing a pressure term 
in the momentum equation (\ref{jeans5}) yielding
\begin{equation}
\frac{\partial \mathbf{u}}{\partial t}
+ \mathbf{u} \cdot\nabla \mathbf{u}
= -\nabla V - \frac{1}{\rho} \nabla P,
\end{equation}
where $P(\rho)$ is the fluid pressure, a local-quantity given by a specified
equation
of state which takes into account velocity
dispersion.\footnote{In 
Appendix \ref{sec_lb}, we consider a closure of the Jeans equations based on a
generalized Fokker-Planck equation \cite{1996ApJ...471..385C} 
obtained from the coarse-graining of the VP equations in the context of the
theory of violent relaxation \cite{1967MNRAS.136..101L}. This parametrization is
then
generalized to the case of bosonic particles described by the SP equations.}
This amounts to closing the
hierarchy
of Jeans equations with the
isotropy ansatz $\mathbb{P}_{\rm
J}=P(\rho) \mathbb{I}$. In this manner, there is no shell-crossing
singularities. The velocity dispersion can be a consequence of the
multi-streaming or it can
be already present in the initial condition. On general
grounds, we expect that
$\mathbb{P}_{\rm
J}\rightarrow 0$ at large scales while it is nonzero at small
scales in order to avoid singularities. Therefore, the limit $\mathbb{P}_{\rm
J}\rightarrow  0$ is different from   $\mathbb{P}_{\rm
J}=0$. This is similar to the remark previously made for the SP
equations where the parameter $\hbar/m$ controls the small-scale
(quantum) resolution. The SP equations in Madelung form can be written as
\begin{equation}
\frac{\partial \mathbf{u}}{\partial t}
+ (\mathbf{u} \cdot\nabla )\mathbf{u}
= -\nabla V - \frac{1}{\rho} \nabla \cdot\mathbb{P}_{\rm Q},
\end{equation}
where from Eq. (\ref{Eq:PQ}):
\begin{equation}
\mathbb{P}_{\rm Q} = \frac{\hbar^2}{4m^2} \left(\frac{1}{\rho} 
\partial_i \rho \partial_j\rho -
\partial_{ij}\rho\right ).
\end{equation}
Interestingly this bares some resemblance to the Jeans stress-tensor, 
with the velocity average operators replaced by density gradients (a non-local
quantity). For
$\hbar/m\rightarrow 0$,   $\mathbb{P}_{\rm
Q}\rightarrow 0$ at large scales while it is nonzero at small scales, thereby 
avoiding singularities. This is qualitatively similar to the
expected behavior of $\mathbb{P}_{\rm
J}$ in the Jeans equations. This is a first hint why the SP
equations should return the VP equations when $\hbar/m\rightarrow 0$. The
quantum pressure tensor acts as a regularizer of caustics and singularities in
classical solutions.

In conclusion, the VP and SP equations are superior to the
pressureless EP equations. They take into account velocity dispersion whereas
the pressureless fluid description does not. They can be used to describe
multistreaming and caustics in the nonlinear regime, whereas the pressureless
fluid equations break down in that regime.

\subsubsection{From wave functions to distribution functions}
Following the reasoning of \cite{1993ApJ...416L..71W} we may 
construct a phase-space representation of any given wave function
$\psi(\mathbf{x},t)$  via the so called Husimi representation, which is
essentially a smoothed version of the Wigner quasiprobability distribution
\citep{husimi1940}. It is given by
\begin{multline}
\label{eq:husimi}
\Psi ( \mathbf{x}, \mathbf{p}, t ; \eta) = \left(\frac{1}{2\pi\hbar} \right)^{n/2} \left(\frac{1}{\pi \eta^2}
 \right)^{n/4}\\
 \times \int \mathrm{d}^n {r} \psi(\mathbf{r},t) \exp 
\left(-\frac{(\mathbf{x}-\mathbf{r})^2}{2 \eta^2} -i\frac{\mathbf{p} \cdot
(\mathbf{r}-\mathbf{x}/2)}{\hbar}\right),
\end{multline}
where $n$ represents the number of dimensions under consideration in the
problem.  Examining the above we can see that the constructed distribution
essentially spatially smoothes out the wave function $\psi$
with a Gaussian window of width $\eta$, as well as 
performing
a sort of Fourier transform to obtain an associated momentum. In order to
construct something similar to a distribution function we then define the
following quantity:
\begin{equation}
\label{eq:schrodinger_dist_func}
\mathcal{F}(\mathbf{x},\mathbf{p},t) \equiv
\lvert\Psi(\mathbf{x},\mathbf{p},t)\rvert^2.
\end{equation}
To illustrate how this representation works, we will apply it later to  the simple case of the harmonic oscillator (Section~\ref{sec:sho}). However, from this definition one can directly show \citep{1989PhRvA..40.2894S}:

\begin{equation}
\label{eq:quant_cbe}
\frac{\partial \mathcal{F}}{\partial t} = \sum_{i=1}^3 \left( m\frac{\partial V}{\partial x_i} \frac{\partial \mathcal{F}}{\partial p_i} - \frac{p_i}{m}\frac{\partial \mathcal{F}}{\partial x_i}\right) + \mathcal{O}(\hbar) + \mathcal{O}(\hbar^2) + \dots
\end{equation}
So we see that, as one might hope, we have an exact analog of the Vlasov equation, plus quantum corrections.

\subsubsection{From distribution functions to wave functions}
\label{sec:ic}

We describe in this subsection how to encode the initial conditions of a classical distribution function 
into the wave function.

In the case that the distribution function is cold/single-stream, 
i.e., a single velocity $\mathbf{u}=\mathbf{v}$ at a given position, 
one may solve for the phase of the wave function by 
solving the Poisson problem:
\begin{equation}
\nabla\cdot\mathbf{u}=\nabla^2 S/m
\label{eqn:ic}
\end{equation}
for the quantum phase  $S$.
The amplitude of the wave function is given by the square-root of the density $\sqrt{\rho}$.
Thus wave function is reconstructed per Eq.~(\ref{eqn:decomp}).

In the case of multi-stream or warm initial conditions, 
the correspondence is a bit more complicated
as the densities will not match exactly but the quantum wave function will include interference patterns.
The idea is described in \cite{1993ApJ...416L..71W}.
The wave function on a discretized grid can be constructed as
\begin{equation}
\psi(\mathbf{x}) \propto \sum_{\mathbf{v}} \sqrt{f(\mathbf{x},\mathbf{v})} {\rm
e}^{i m \mathbf{x}\cdot\mathbf{v}/\hbar + 2\pi i \phi_{{\rm rand},\mathbf{v}}}
\, d^3v
\label{eqn:init}
\end{equation}
with normalization such that the total mass 
in the classical and quantum cases are the same $M=\int\int f
\, d^3v d^3x = \int \lvert\psi\rvert^2\, d^3x$.
Here the sum is over the discretely sampled velocities $\mathbf{v}$ and $\phi_{{\rm rand},\mathbf{v}}\in[0,2\pi)$ is a random phase associated with velocity $\mathbf{v}$, necessary so that the different phases associated with a given fluid velocity are uncorrelated.

Note the reverse -- constructing a distribution function from a wave function -- may be done with Husimi smoothing \citep{1993ApJ...416L..71W}.
However, some wave functions
may not have a classical equivalent, 
e.g. discontinuities on the wave function phase
are valid in a quantum system but correspond to infinite classical velocities (it is important to note that this can only happen at regions where the density approaches zero).

\section{Numerical Spectral Method}
\label{sec:method}

The SP equations are evolved numerically using the unitary spectral method developed in \cite{2017MNRAS.471.4559M}.
The method uses a 2nd-order time-stepping method and gives exponential convergence in space. The timesteps are decomposed into a kick-drift-kick symplectic leapfrog-like scheme, where each `kick' and `drift' are unitary operators acting on the wave function.

The spectral method proves to be useful at capturing 
vortex lines/cores, which turns out to be an integral feature
of cosmological Bose-Einstein condensate halos \citep{2017MNRAS.471.4559M}. 
Vortex lines are locations where the density $\rho=0$, but are sites of quantized vorticity in the fluid (the rest of the fluid is vortex free, as $\nabla\times\mathbf{u}=0$ since the velocity is the gradient of a scalar (the phase) per Eq.~(\ref{eqn:vmadel})).
An alternative approach would be to work in the fluid (Madelung) formulation, and use fluid solver methods such as the smooth-particle-hydrodynamics approach
developed in \cite{2015PhRvE..91e3304M}, but the 
fluid formulation may prove difficult to capture vortex lines accurately, as the density is $0$ and the velocity is formally infinite here.
The spectral method, with its exponential spatial convergence, and keeping track of the phase directly rather than the velocity, is able to handle vortex lines without difficulty.

The spectral method requires a uniformly gridded domain.
As the velocity is defined as the gradient of the phase, 
the spatial resolution defines a maximum
velocity that can be numerically represented:
\begin{equation}
v_{\rm max}=\frac{\hbar}{m}\frac{\pi}{\Delta x}.
\label{eqn:grad}
\end{equation}
This sets a resolution requirement for our simulations.

The VP equations are evolved using a standard second-order symplectic particle
mesh solver in 1D and 2D \citep{2017MNRAS.465.3154M}, and a  hybrid tree-nested particle mesh approach in 3D using the $N$-body gravitational solver of the {\sc Arepo} code \citep{2010MNRAS.401..791S}.

\section{Numerical Simulations and Results}
\label{sec:sims}

We carry out several problems in 1D, 2D, and 3D to demonstrate 
the SP--VP correspondence.

\subsection{1D fixed potential} \label{sec:sho}
First, let us consider a simple 1D case, in which the potential is fixed, in
order to learn about the regularization role the quantum pressure tensor has on the 
classical solution.
The 1D simulations presented here are meant to help form intuition about the
problem, and later, more complicated simulations in 2D and 3D with the self-potential, are more rigorously
analyzed.
1D simulations with self-gravity, as a means to recover the classical limit, have also been extensively simulated previously
\citep{1993ApJ...416L..71W}, which is why we will focus on 2D and 3D simulations with self-gravity in
sections~\ref{sec:2da}, \ref{sec:2db}, \ref{sec:2dc}, and \ref{sec:3d}, where
additional quantum mechanical structures, such as vortices, 
introduce new non-trivialities in the correspondence.

\begin{figure}
\includegraphics[width=0.47\textwidth]{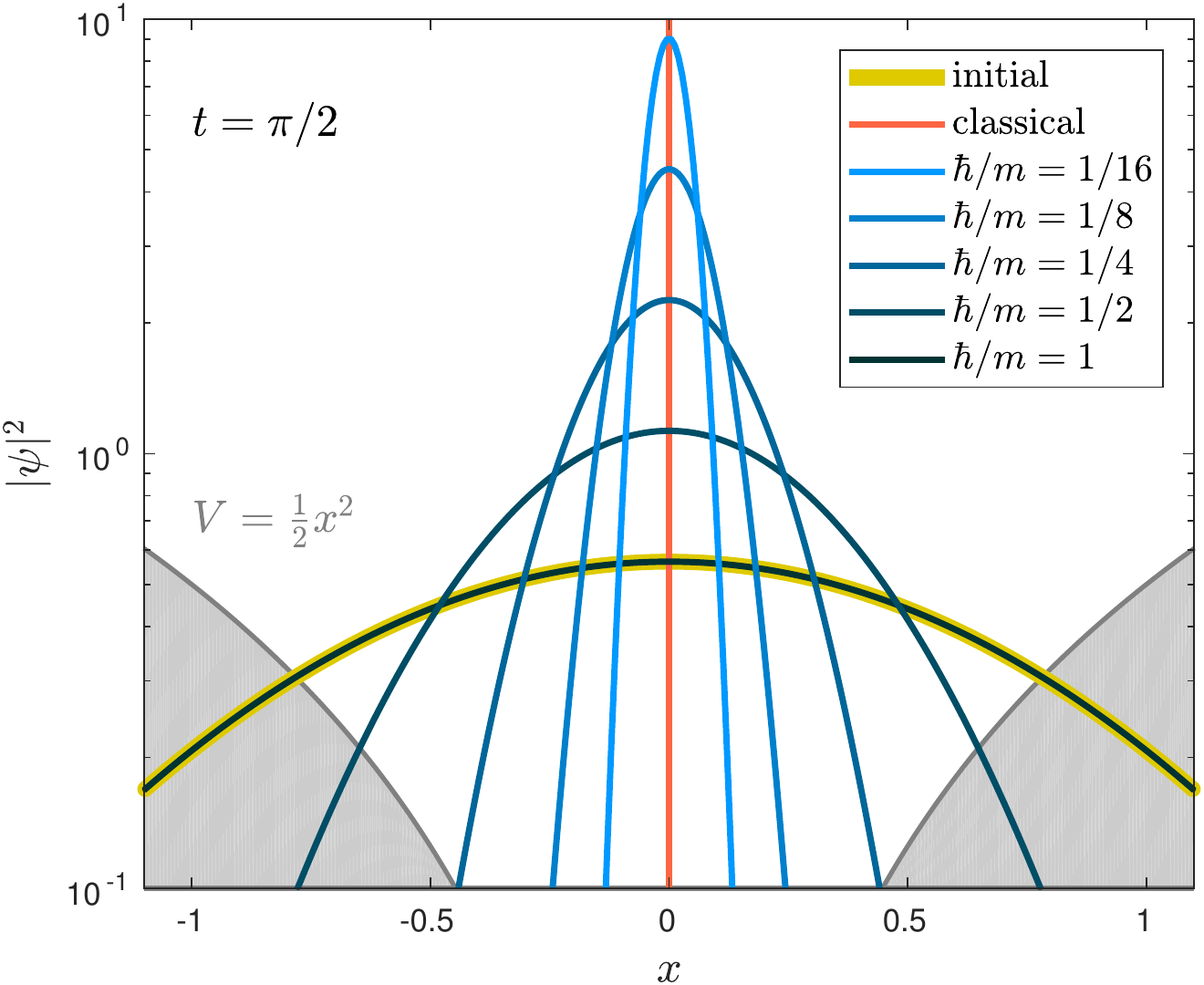}
\caption{Oscillatory motion of a wave function/collisionless particles in a simple harmonic potential. At $t=\pi/2$ the initial Gaussian distribution collapses into a $\delta$-function under the VP equations, which is captured as a Gaussian of width $\hbar/m$ by the SP equations.}
\label{fig:sho1d}
\end{figure}

\begin{figure}
\includegraphics[width=0.47\textwidth]{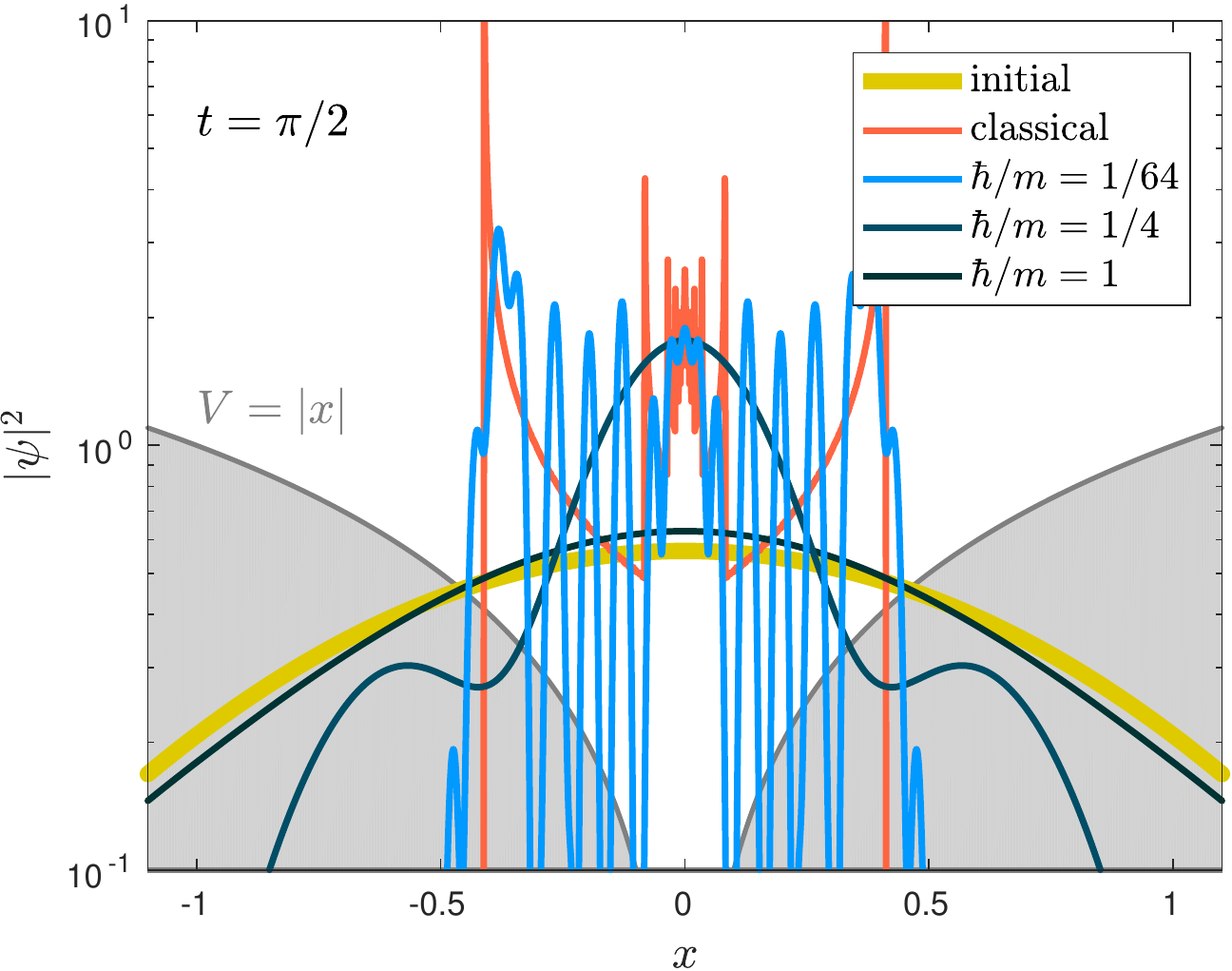}
\caption{In a linear potential, an initially smooth  Gaussian distribution function develops caustics via the VP equations. These are approximated by a superposition of Airy function solutions in the SP equations to a greater and greater degree as $\hbar/m\to 0$. The solution is plotted at $t=\pi/2$.}
\label{fig:lin1d}
\end{figure}

\subsubsection{Harmonic potential}

The setup of the simple problem is as follows.
Consider a simple harmonic oscillator potential
$V=\frac{1}{2}x^2$.
Initially, the wave function is a Gaussian
\begin{equation}
\psi = {\rm e}^{-x^2/2}\pi^{-1/4}
\end{equation}
which is the ground state of the system 
when $\hbar/m=1$.
As such, the system would remain time-independent.
In the classical analog, 
the density is initially $\rho=\lvert\psi\rvert^2$
and the velocity is $0$.
Here, individual collisionless particles
all undergo sinusoidal motion 
with the same period $T=2\pi$.
As such, at time $t=\pi/2$
all the particles reach $x=0$, 
and hence the density becomes a Dirac delta function $\delta(x)$ -- a simple example of a caustic. The particles return to their initial state at time $T$.
We are interested in the behavior of the quantum system as we send $\hbar/m\to 0$.

This case has an analytical solution which one can derive from the Feynman
propagator for this system (see, for example,
\cite{2003AmJPh..71..483B}). One can write the
wave function with these initial conditions as a function of space and time as:
\begin{multline}
\label{eq:1d_sho}
\psi(x,t) = \pi^{-1/4} \left(i \frac{\hbar}{m}\sin(t) + \cos(t) \right)^{-1/2}\\
\times \exp \left[\frac{-2x^2 +i\sin(2t)(\frac{\hbar}{m} -
\frac{m}{\hbar})}{4\left(\cos^2(t) + \left(\frac{\hbar}{m} \right)^2 \sin^2(t)
\right)} \right].
\end{multline}
We see, in accordance with Figure \ref{fig:sho1d} that when $ t = \frac{\pi}{2}
$ we have the density $\rho \equiv \lvert \psi \rvert^2  $ as:
\begin{equation}
\label{eq:rho_sho}
\rho\left(x,t=\frac{\pi}{2}\right) =
\frac{m}{\hbar\sqrt{\pi}} \exp\left[- \left(\frac{x
m}{\hbar} \right)^2 \right]
\end{equation}
which is simply a Gaussian of width $\sigma =2^{-1/2}\hbar/m$.
Thus, what would classically result in a caustic $\delta$-function under the
evolution of the Vlasov equation, is `regulated' here by a width which is
linearly proportional to the `quantumness' of the system: $\hbar/m$.

For this simple case we can also construct the Husimi
representation using Eq. (\ref{eq:husimi}).
In order to compare this representation with Figure \ref{fig:sho1d}, we can
examine what $\mathcal{F}_{\rm SHO}\left(x,p,t=\frac{\pi}{2}\right)$ would look
like in this example. Computing this we find:
\begin{multline}
\label{eq:sho_husimi}
\mathcal{F}_{\rm SHO}\left(x,p,t=\frac{\pi}{2} \right)
 = \frac{\sqrt{2}m \eta }{\pi(\hbar^2 + 2 m^2 \eta^2)}\\
\times\exp\left[-\frac{p^2\eta^2 + 2 x^2 m^2}{\hbar^2 + 2 m^2 \eta^2} \right].
\end{multline}
Compare this with the classical collisionless solution:
\begin{equation}
\label{eq:sho_classical}
\mathcal{F}_{\rm classical}\left(x,p,t=\frac{\pi}{2} \right) = 
\frac{1}{\sqrt{2 \pi}m} \exp\left[-\frac{p^2}{2 m^2} \right] \delta(x)
\end{equation}
which is the limit of Eq.~(\ref{eq:sho_husimi}) as $\hbar\to 0$ and $\eta\to 0$. 

We simulated the problem with $N=1024$ grid points on the domain $[-4,4]$.
The solution is sinusoidal and
at time $t=\pi/2$ the wave function is a Gaussian 
with a narrowed width of $\sigma = (\hbar/m)$, 
as shown in Figure~\ref{fig:sho1d}.
Hence, the solution approaches
the classical $\delta$-function solution as $\hbar/m\to 0$.
The infinite is regularized by the quantum uncertainty
principle.

\subsubsection{Linear potential}

Next we consider a slightly more complicated example
by changing the potential to a linear potential
$V=\lvert x \rvert$.
Now, the classical behavior under the VP equation
forms many caustics because the particles have non-synchronized periods.
Individual particles travel in oscillatory fashion with velocity $v=-{\rm sign}(x)$.

In the quantum version of this problem, it is well-known that 
the energy eigenstates are Airy functions \citep{2016iqm..book.....G}. 
The solution at $t=\pi/2$ is plotted in
Figure~\ref{fig:lin1d} for several values of $\hbar/m$, 
where we see that the superposition of Airy functions captures
the
caustics with improved accuracy for smaller values of $\hbar/m$.
At $\hbar/m=1$, the de Broglie wavelength is large and smoothes
out all structures. But as $\hbar/m$ decreases, sharp features are
recovered.

\subsubsection{Discussion}

The 1D examples highlight how the uncertainty principle
regularizes caustics 
and how the classical solution is approached as $\hbar/m\to 0$.
The norm of wave functions approaches the classical limit but
retains order unity oscillations (with period the de Broglie
wavelength) about the classical solution.
Of interest, in the next section, is whether the self-potential from this oscillatory
density field recovers the classical limit.

\subsection{2D cold initial conditions}\label{sec:2da}

\subsubsection{Setup}

\begin{figure}
\begin{center}
\includegraphics[width=0.4\textwidth]{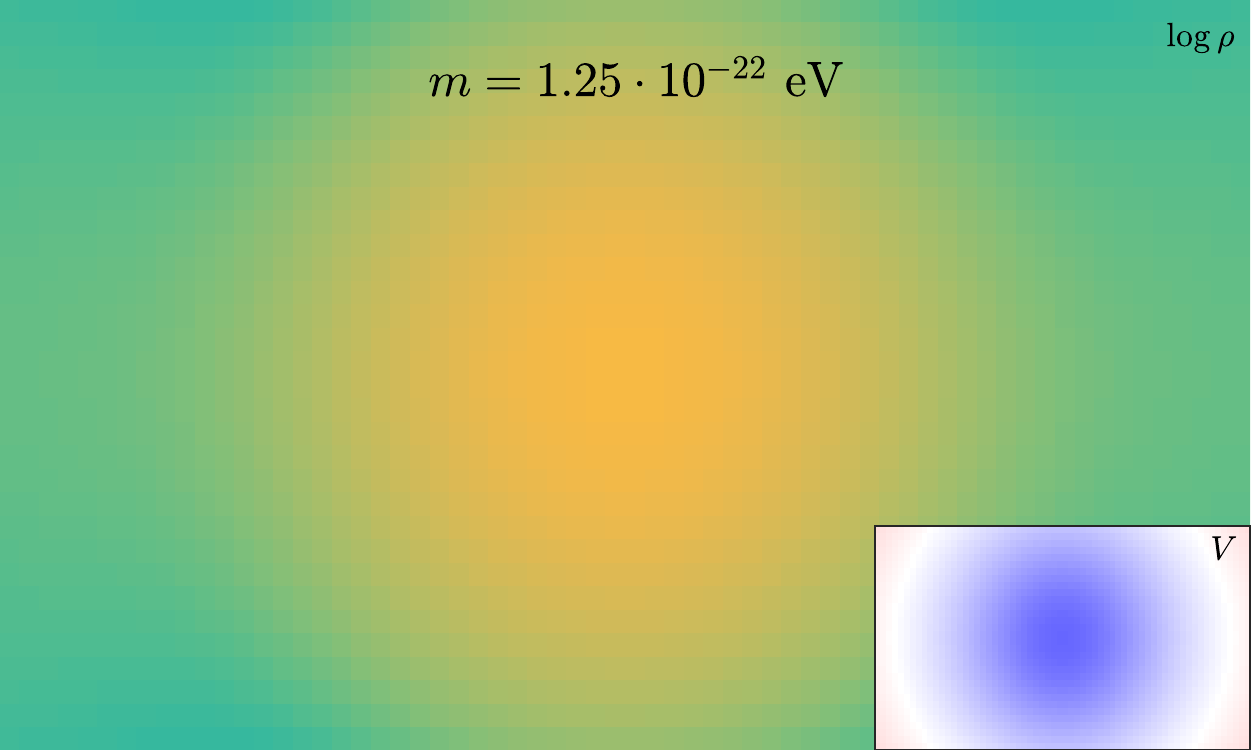}\\
\includegraphics[width=0.4\textwidth]{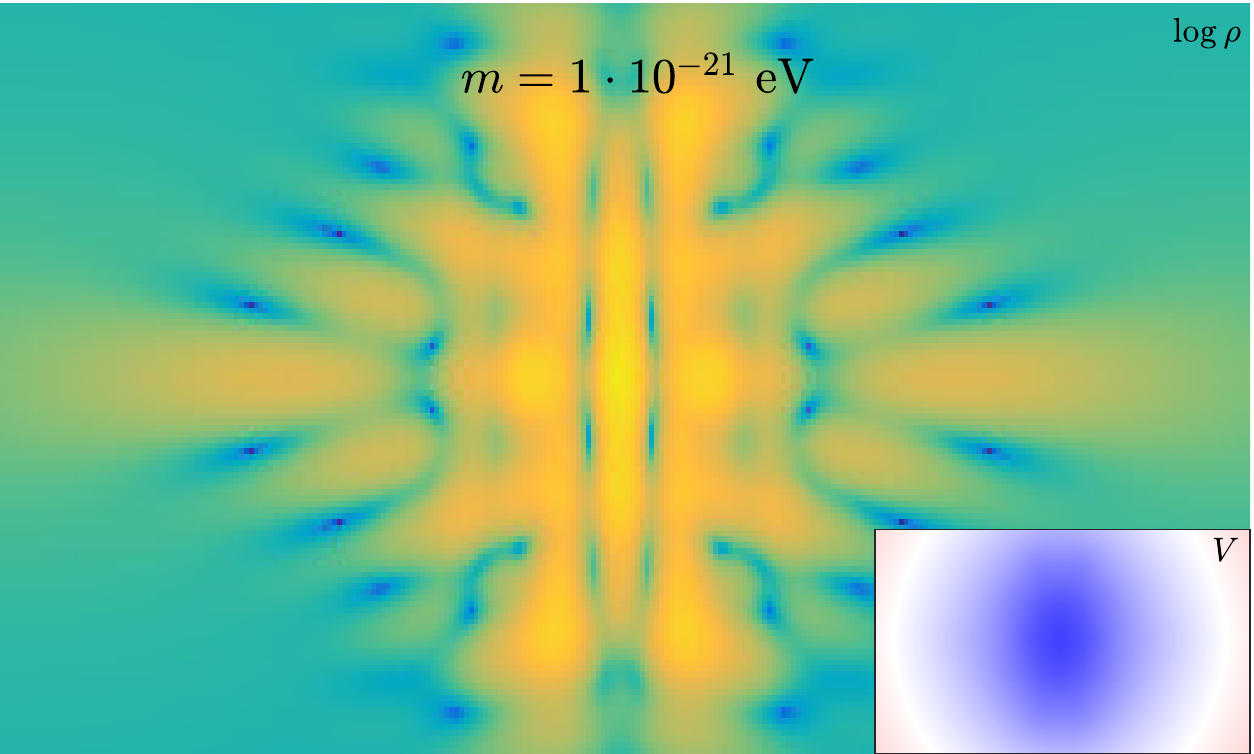}\\
\includegraphics[width=0.4\textwidth]{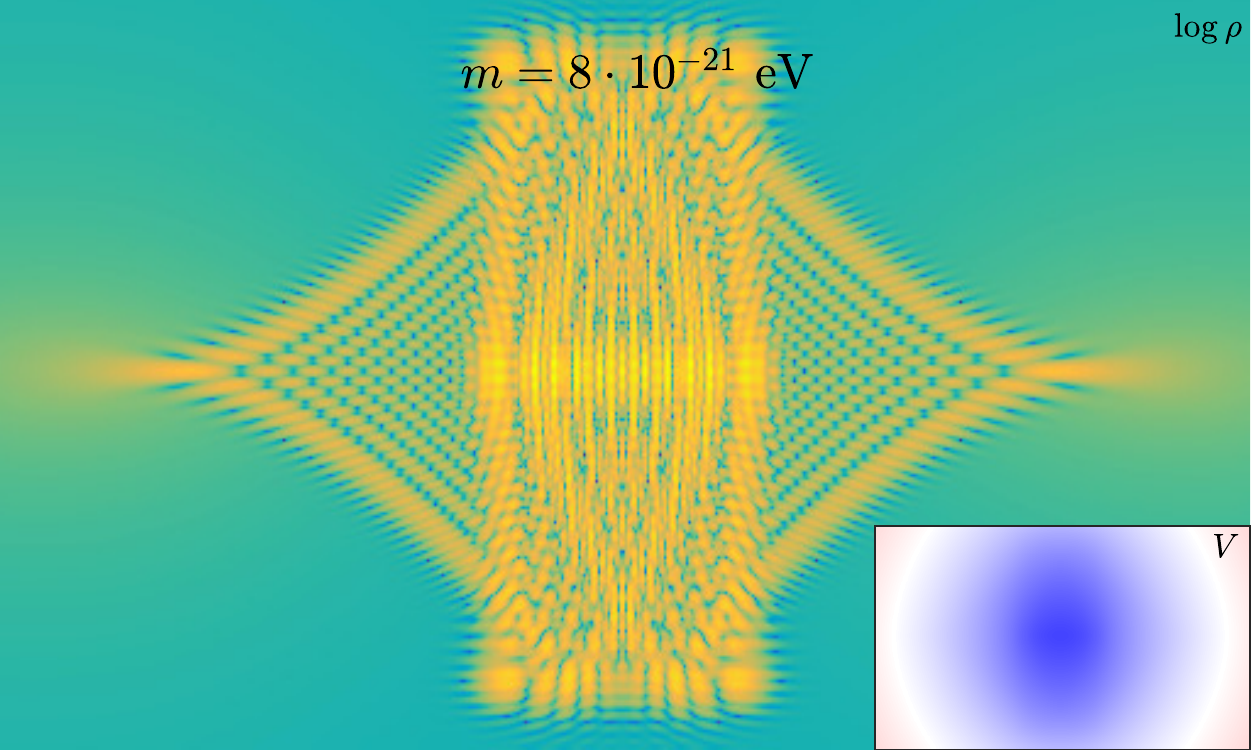}\\
\includegraphics[width=0.4\textwidth]{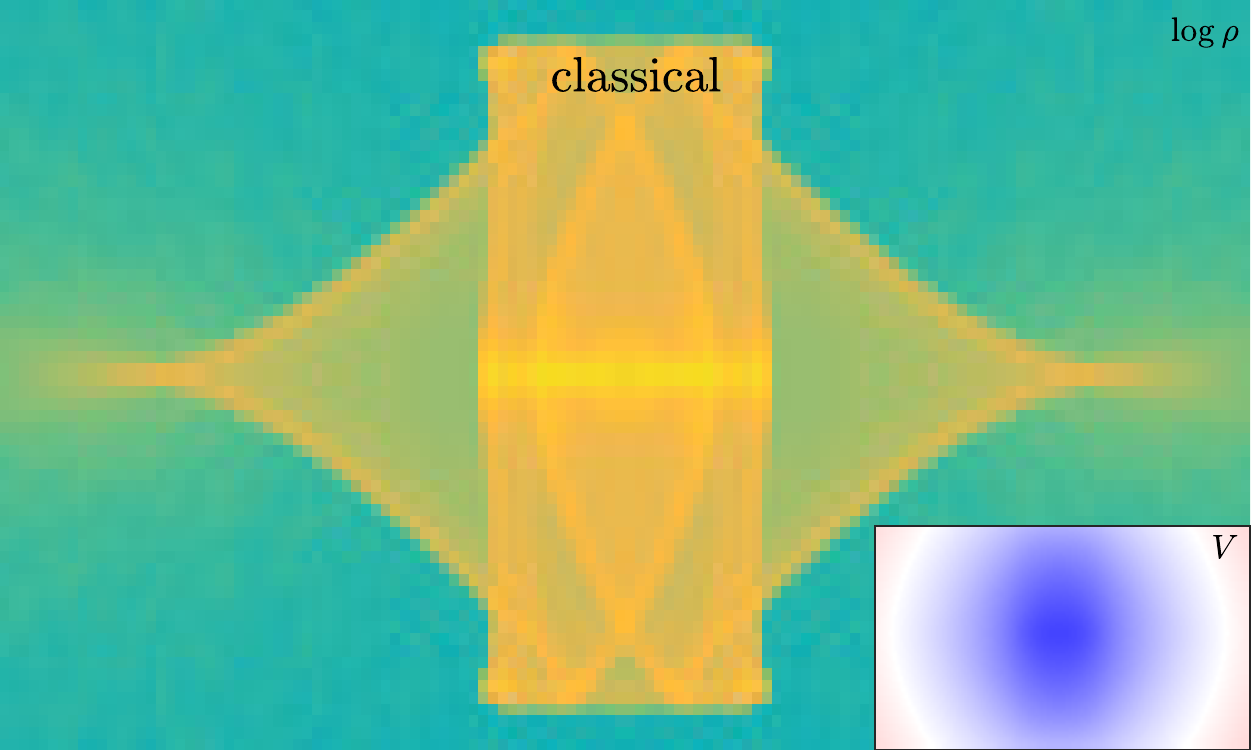}\\
\includegraphics[height=0.22\textwidth]{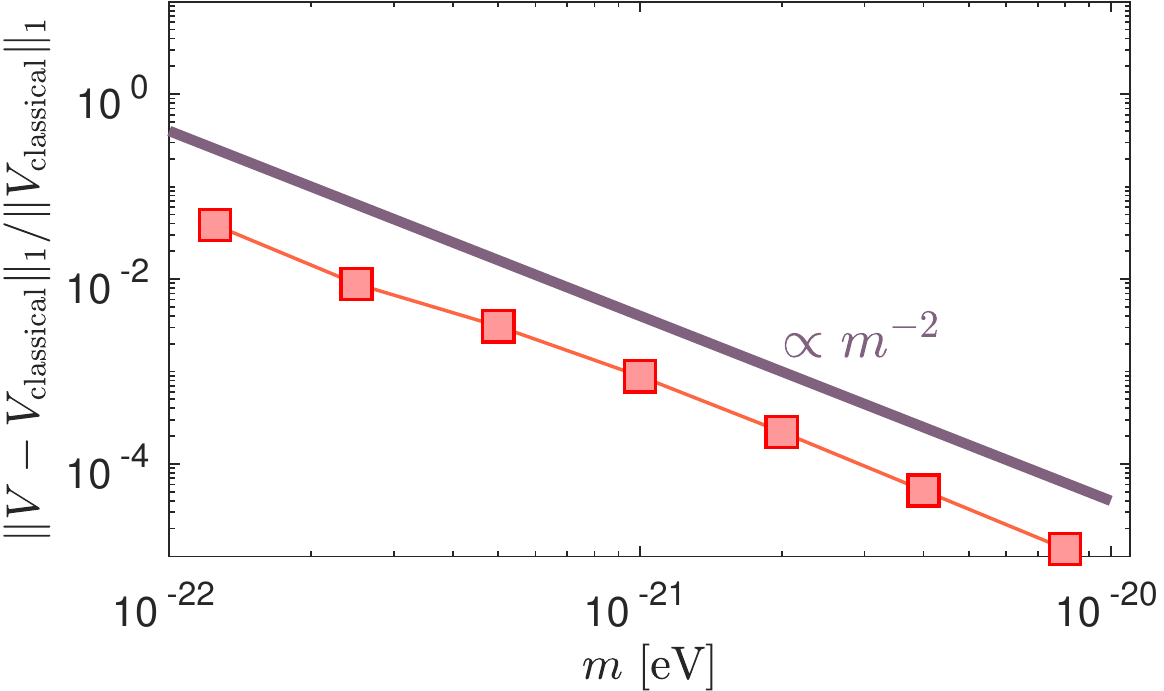}
\includegraphics[height=0.22\textwidth]{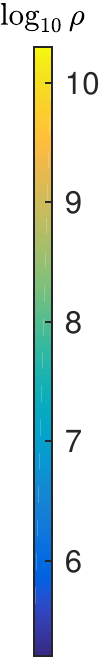}
\includegraphics[height=0.22\textwidth]{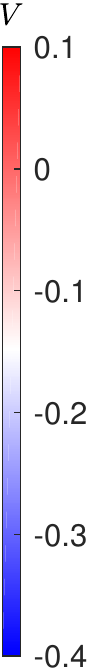}
\end{center}
\caption{
Convergence of the 2D Gaussian collapse problem with cold initial conditions 
to the classical limit. Solution is zoomed in on $[\frac{3}{8},\frac{5}{8}] \times [\frac{7}{16},\frac{9}{16}]$.
The potential converges to the classical limit as $m^{-2}$.}
\label{fig:a2d}
\end{figure}

We consider the collapse 
of two self-gravitating Gaussians
with a 2D simulation.
This can be thought of as the collapse of two parallel cylindrical structures in 3D, 
for example filaments in cosmic structure.
The domain is a periodic box with a side of one Megaparsec (Mpc).
The simulation uses code units of $[L]={\rm Mpc}$, $[v]={\rm km}~{\rm s}^{-1}$, $[M]=M_\odot$.
The initial density is
given by a constant background plus two 2D Gaussians
\begin{equation}
\rho_0 = 
A\left[
\frac{1}{4}
+
{e}^{-\frac{\left(x-\frac{5}{8}\right)^2+\left(y-\frac{1}{2}\right)^2}{
2\sigma^2}} 
+
{e}^{-\frac{\left(x-\frac{3}{8}\right)^2+\left(y-\frac{1}{2}\right)^2}{
2\sigma^2}} 
\right]
\end{equation}
with $A=10^8~M_\odot{\rm Mpc}^{-3}$ and $\sigma=0.1~{\rm Mpc}$. In this cold setup, we assume the initial velocity is $0$,
and hence set the initial phase of the wave function to $0$. We simulate the classical case as well
as the quantum case with 
boson masses ranging between 
$m=1.25\times 10^{-22}~{\rm eV}$
and $m=8\times 10^{-21}~{\rm eV}$
at resolution $2048^2$.

Figure~\ref{fig:a2d} shows a zoom-in on the complicated
structure that forms in the density at time $t=1~{\rm Mpc}~({\rm km}~{\rm s}^{-1})^{-1}$.
The classical solution shows caustic/shell-crossing structure.
We analyze the convergence of the potential $V$ to the 
classical limit under the L1 norm as a function of the boson
mass.

\subsubsection{Discussion}

Importantly, the potential is found to converge to the classical limit as
$m^{-2}$, even though the density profile shows order unity 
oscillations on the scale of the local de Broglie wavelength. In the limit of
large boson mass, the density field recovers the classical limit, except for
interference patterns due to the multiple phase-sheets traveling at different
velocities.
At small boson mass, the entire solution is smoothed out by the uncertainty
principle and the potential is less shallow than the classical
limit since a fraction of the energy of the system is in the quantum gradient
kinetic energy \citep{2017MNRAS.471.4559M}.

\subsection{2D multi-sheet initial conditions}\label{sec:2db}

\begin{figure}
\begin{center}
\includegraphics[width=0.4\textwidth]{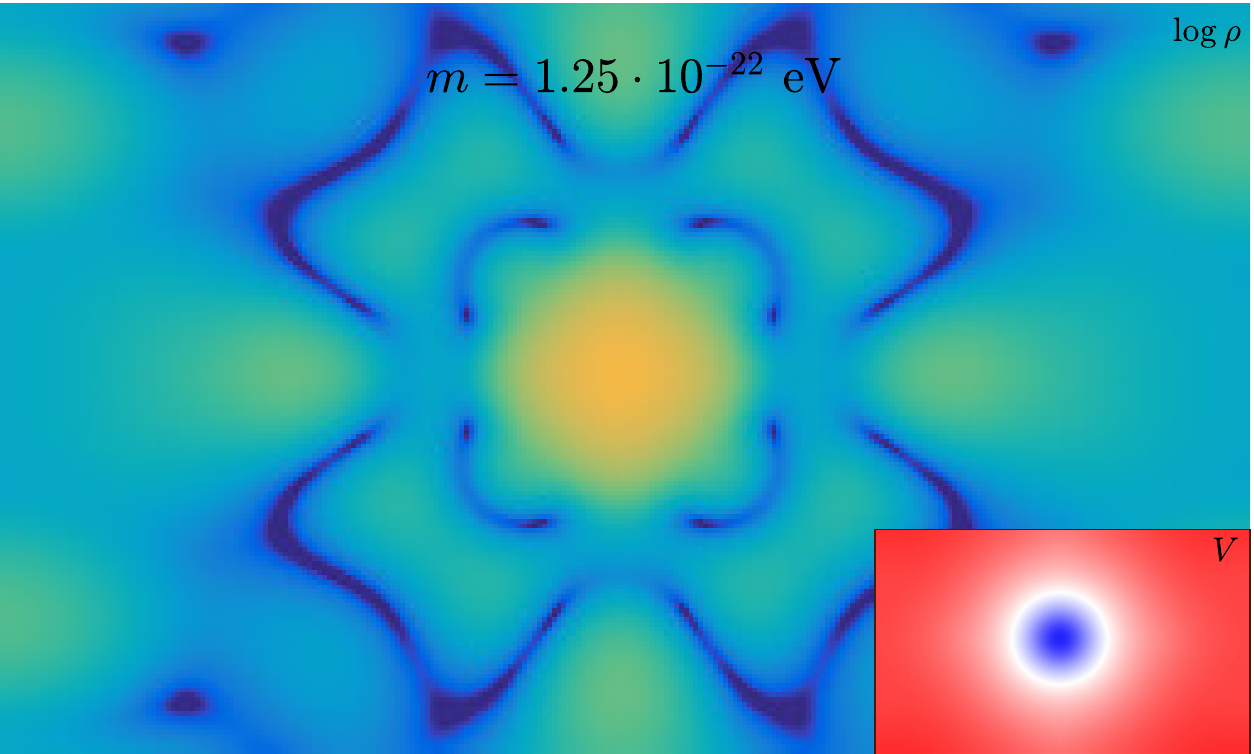}\\
\includegraphics[width=0.4\textwidth]{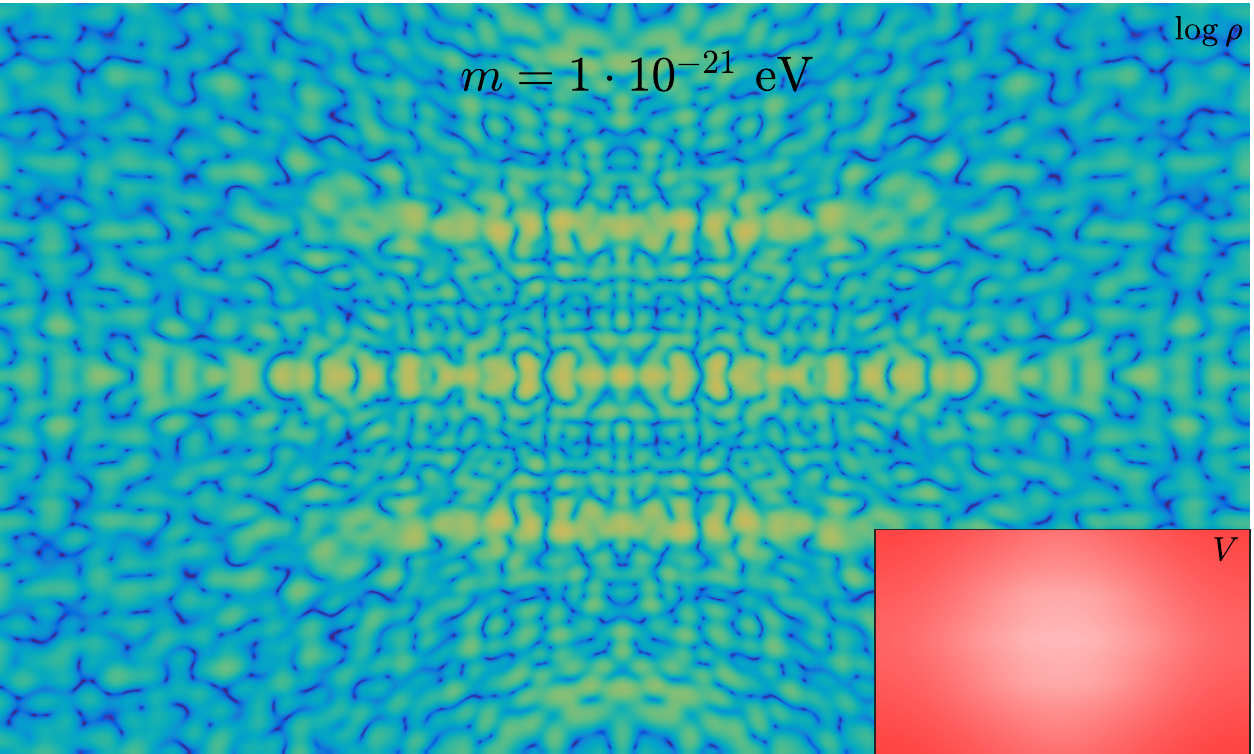}\\
\includegraphics[width=0.4\textwidth]{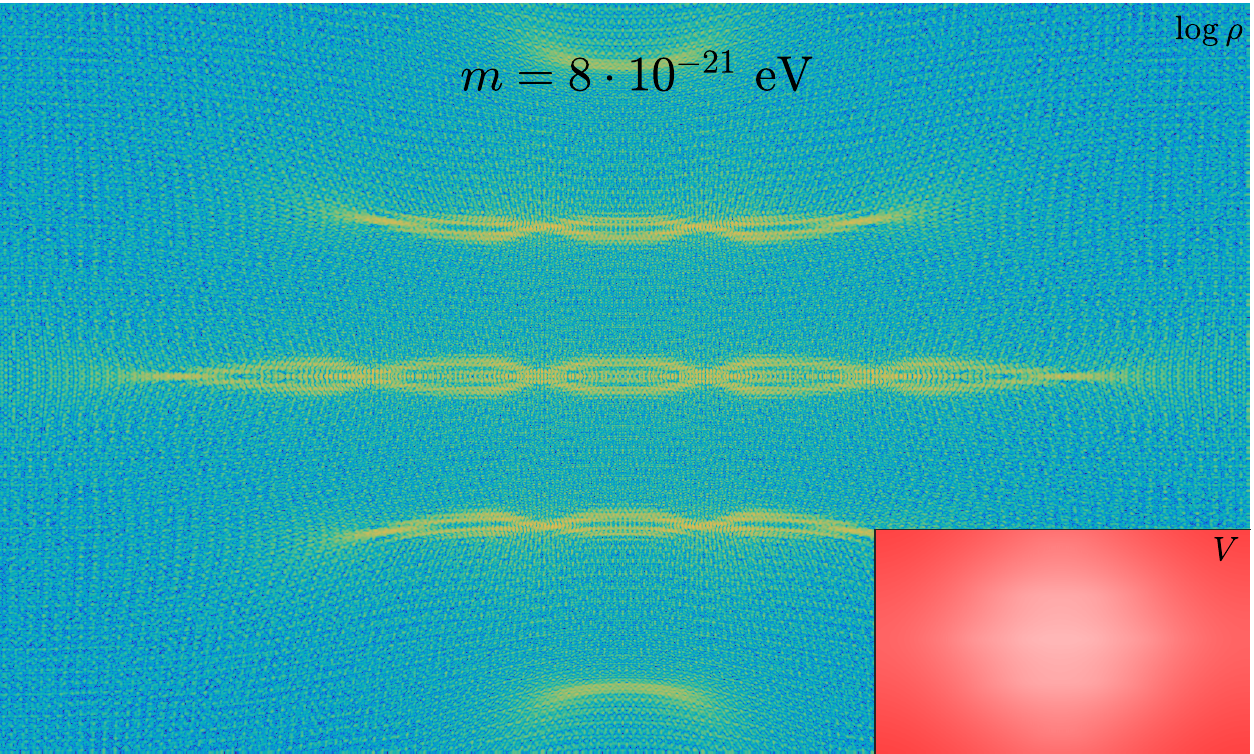}\\
\includegraphics[width=0.4\textwidth]{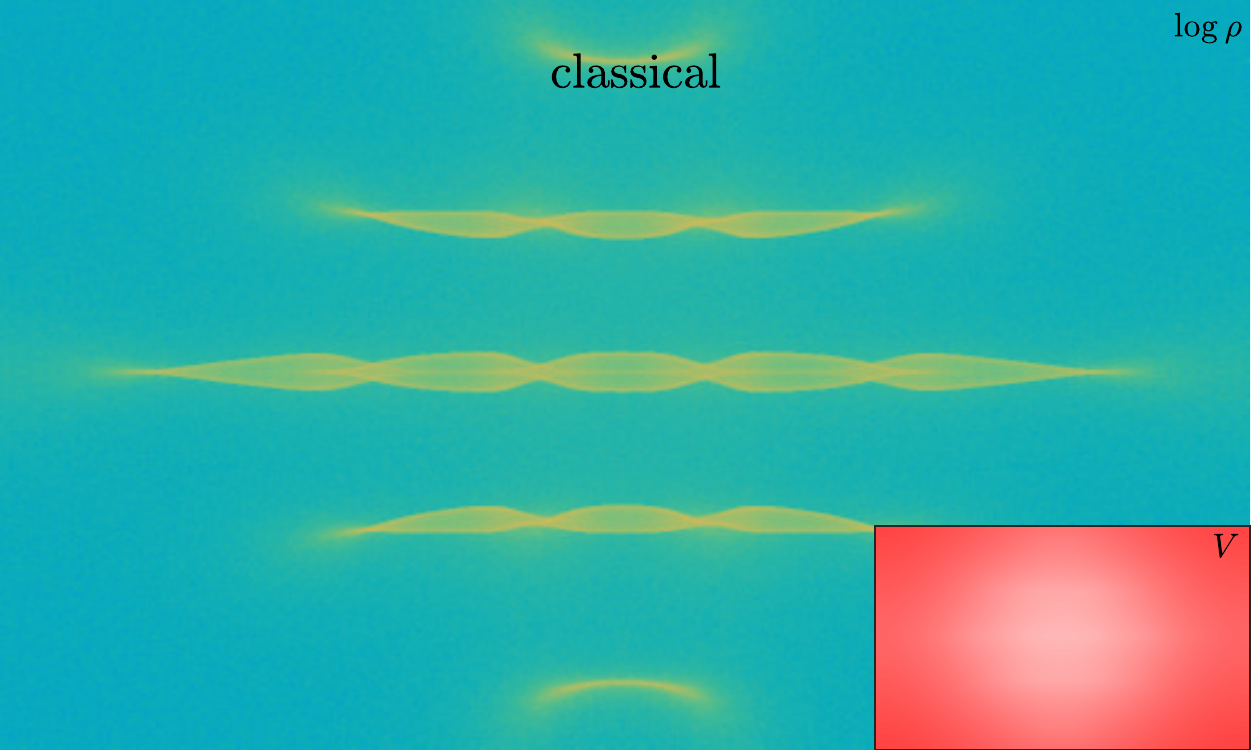}\\
\includegraphics[height=0.22\textwidth]{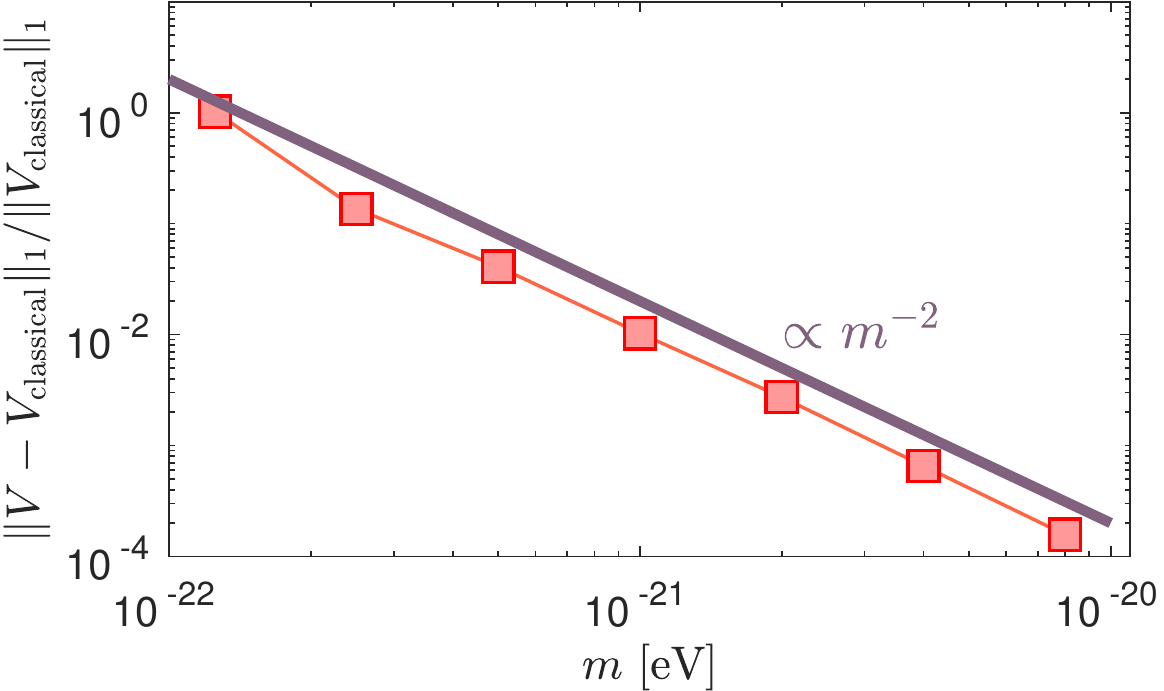}
\includegraphics[height=0.22\textwidth]{cb-eps-converted-to.pdf}
\includegraphics[height=0.22\textwidth]{cb2-eps-converted-to.pdf}
\end{center}
\caption{Convergence of the 2D Gaussian collapse problem with multiple velocity initial conditions 
to the classical limit. Solution is zoomed in on $[0,1] \times [\frac{3}{16},\frac{13}{16}]$.
The potential converges to the classical limit as $m^{-2}$.}
\label{fig:b2d}
\end{figure}

\begin{figure}
\begin{center}
\includegraphics[width=0.4\textwidth]{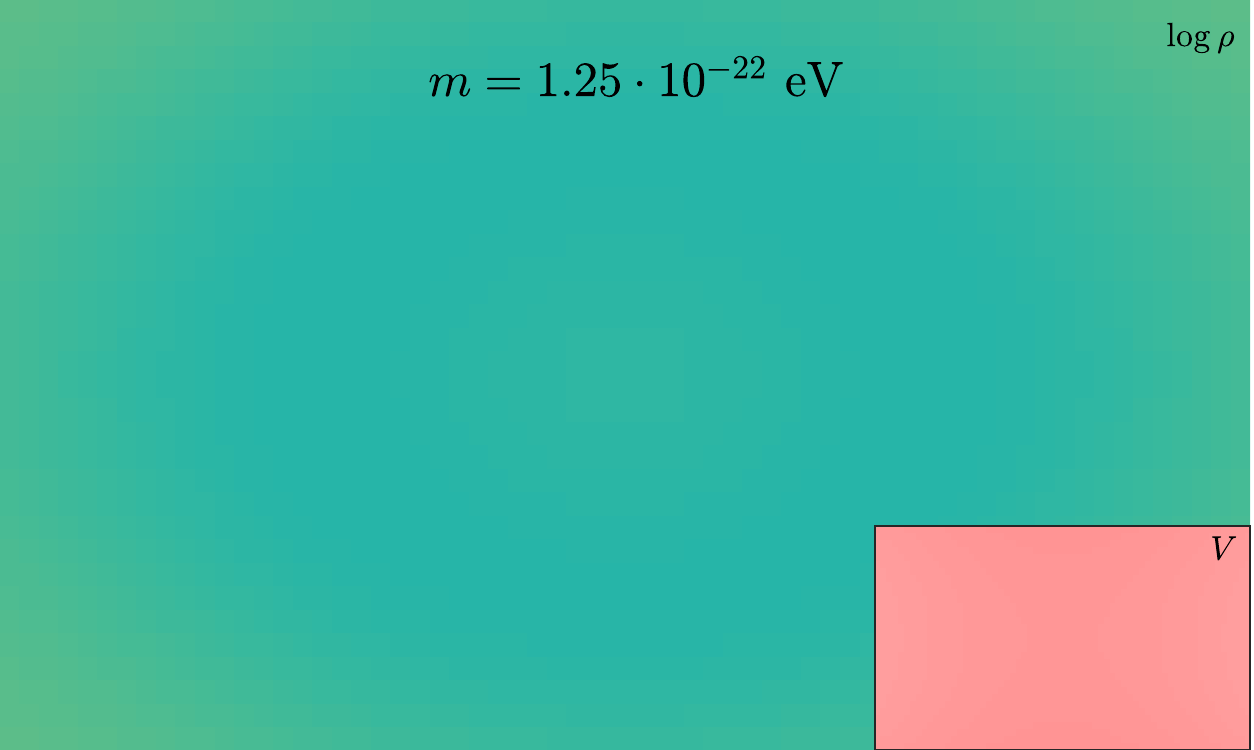}\\
\includegraphics[width=0.4\textwidth]{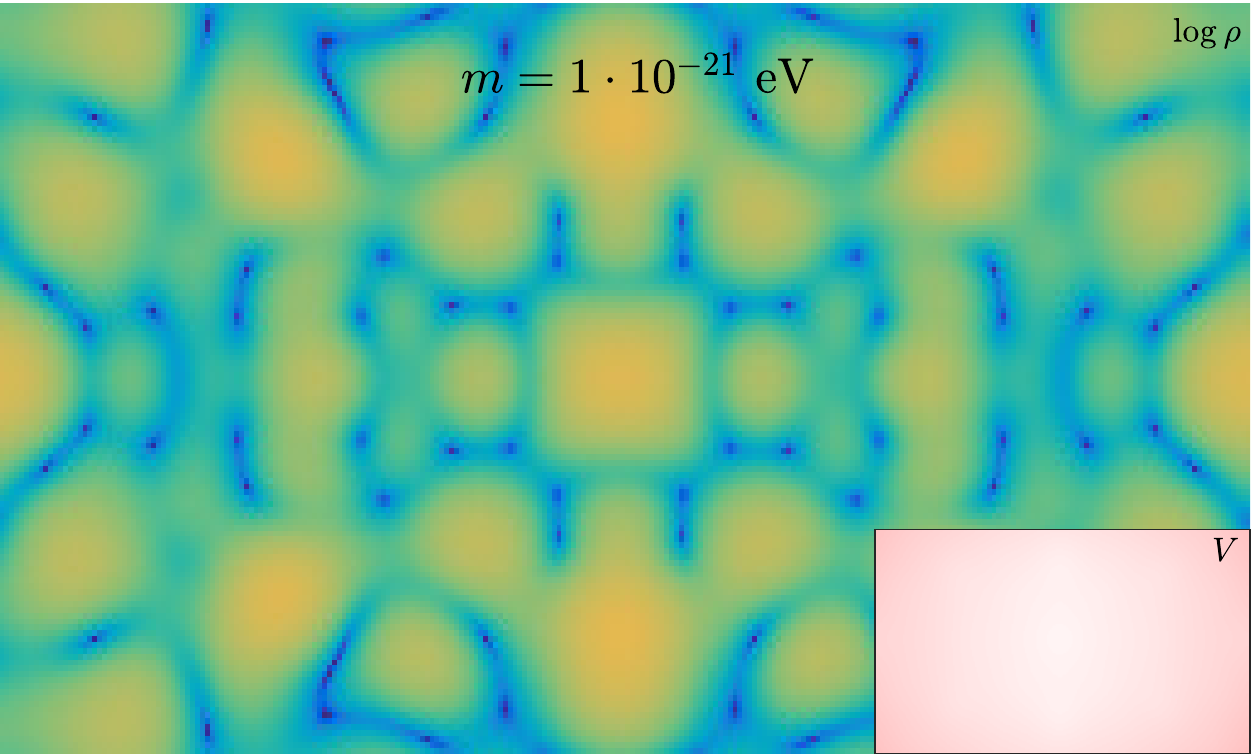}\\
\includegraphics[width=0.4\textwidth]{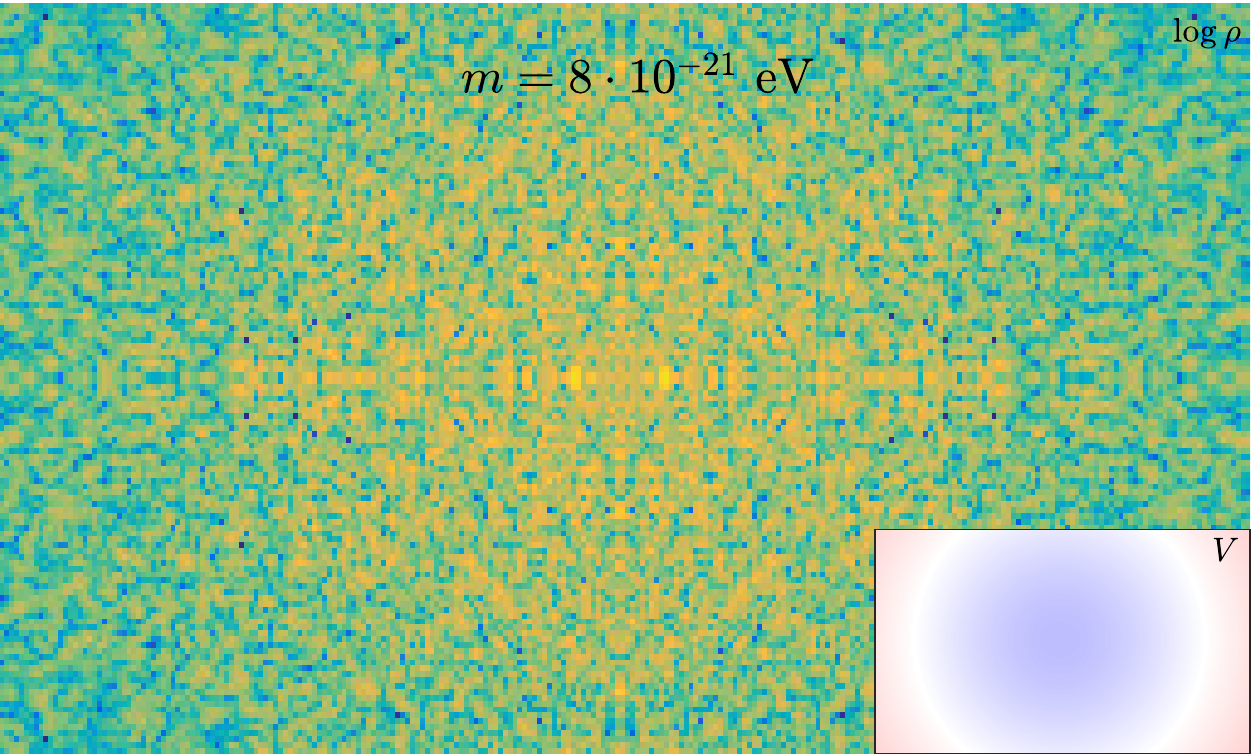}\\
\includegraphics[width=0.4\textwidth]{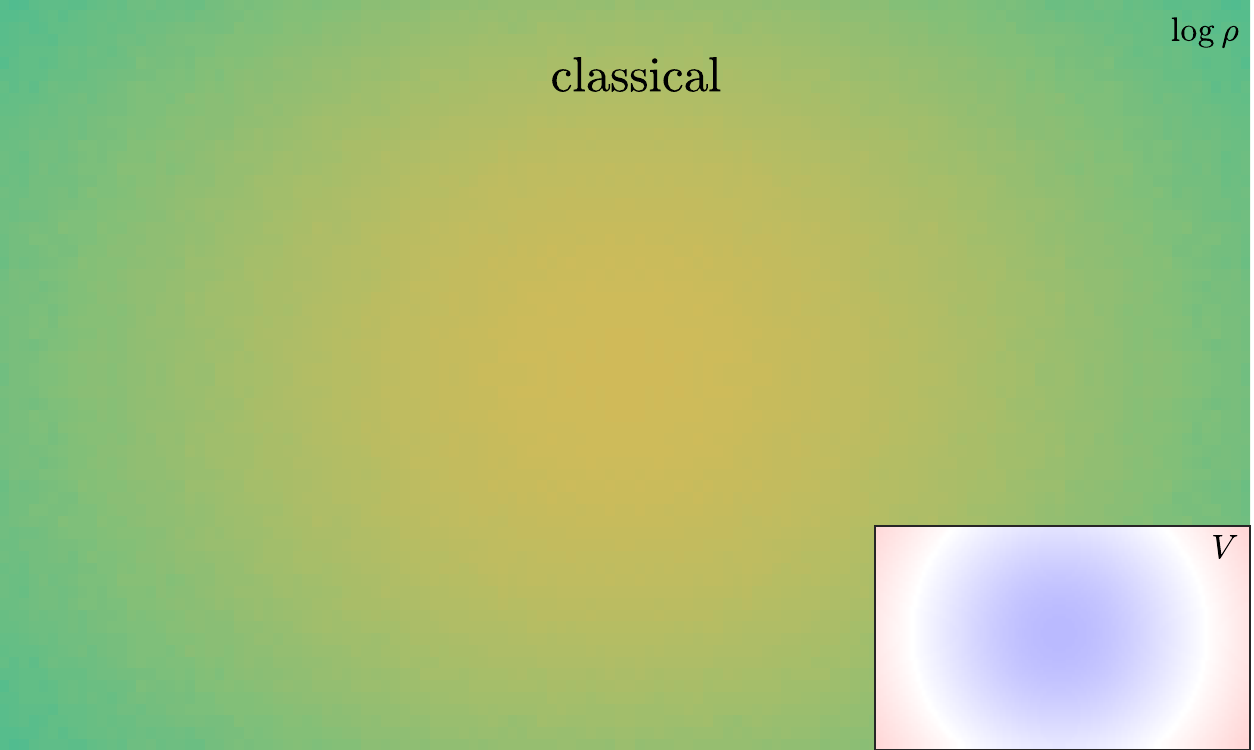}\\
\includegraphics[height=0.22\textwidth]{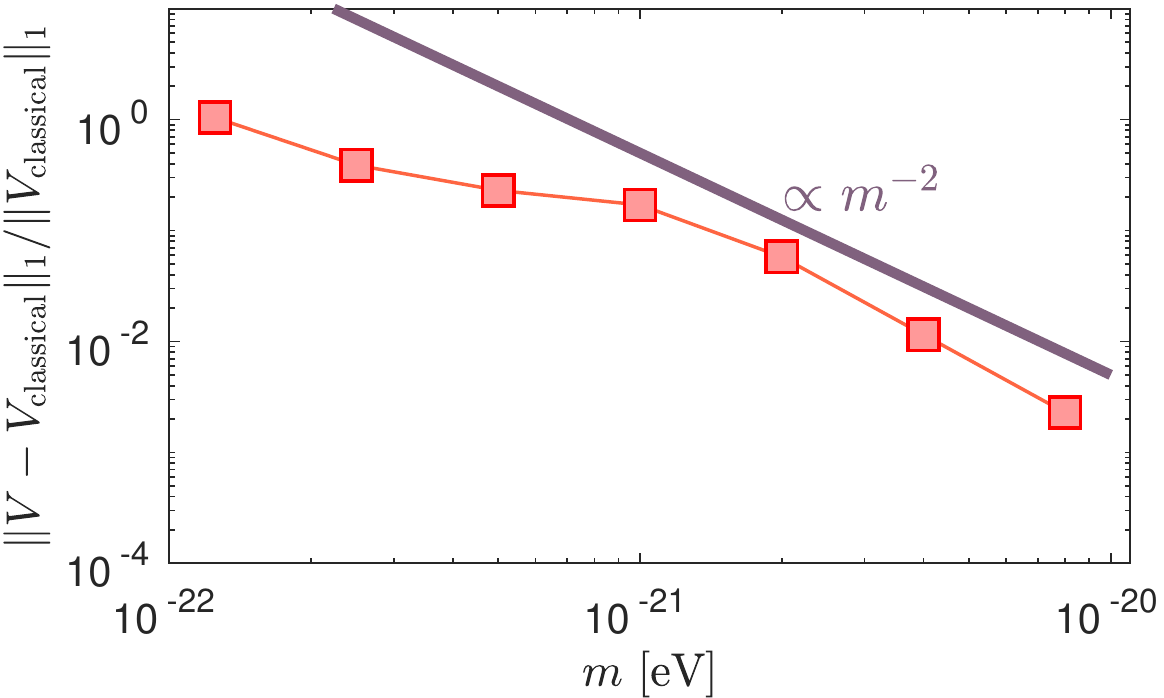}
\includegraphics[height=0.22\textwidth]{cb-eps-converted-to.pdf}
\includegraphics[height=0.22\textwidth]{cb2-eps-converted-to.pdf}
\end{center}
\caption{Convergence of the 2D Gaussian collapse problem with warm initial conditions 
to the classical limit. Solution is zoomed in on $[\frac{3}{8},\frac{5}{8}] \times [\frac{7}{16},\frac{9}{16}]$.
The potential converges to the classical limit as $m^{-2}$.}
\label{fig:c2d}
\end{figure}

\begin{figure*}
\begin{center}
\includegraphics[width=0.42\textwidth]{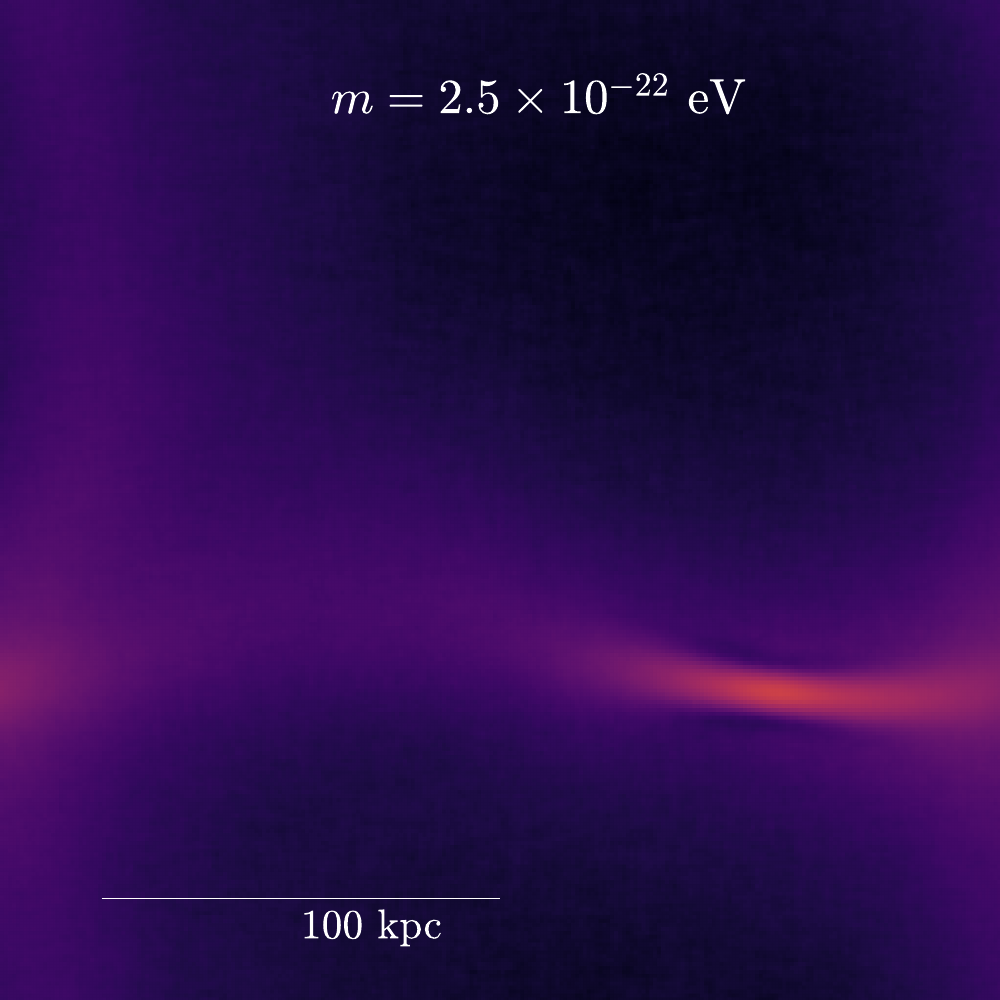}
\includegraphics[width=0.42\textwidth]{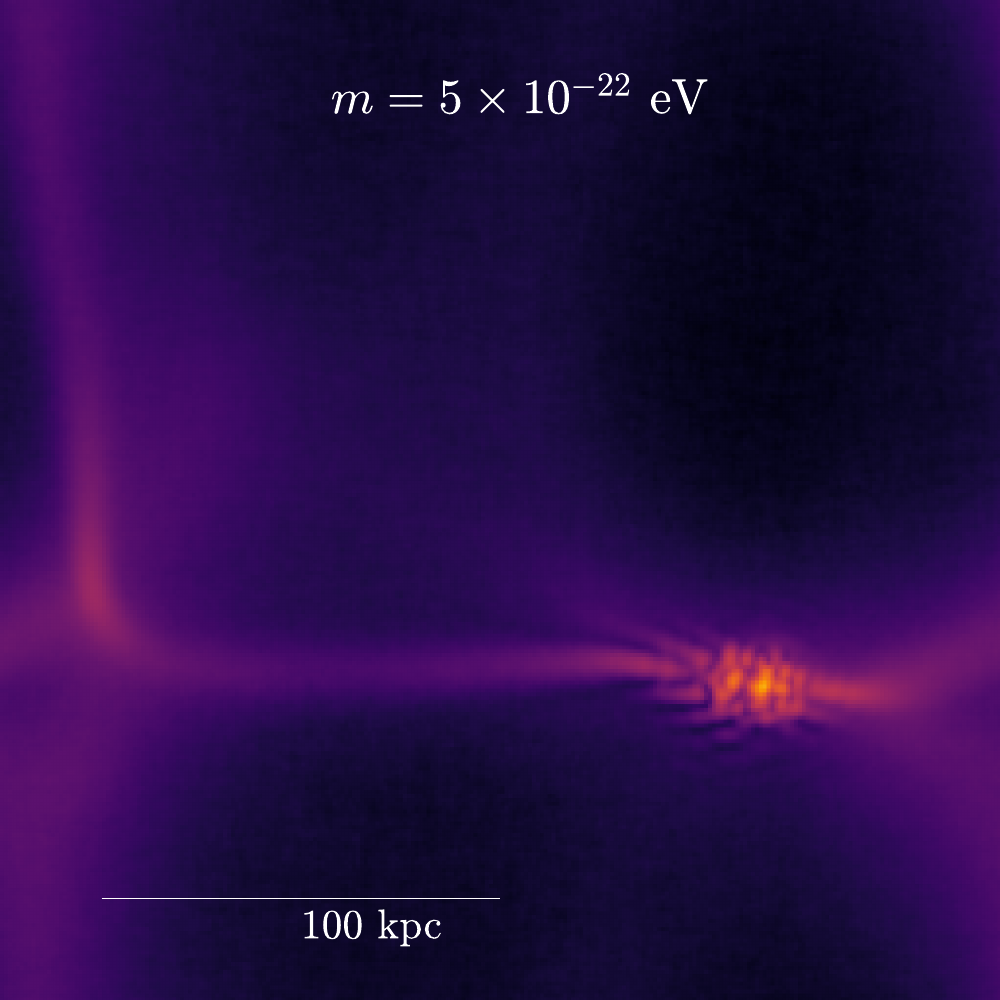}
\includegraphics[width=0.42\textwidth]{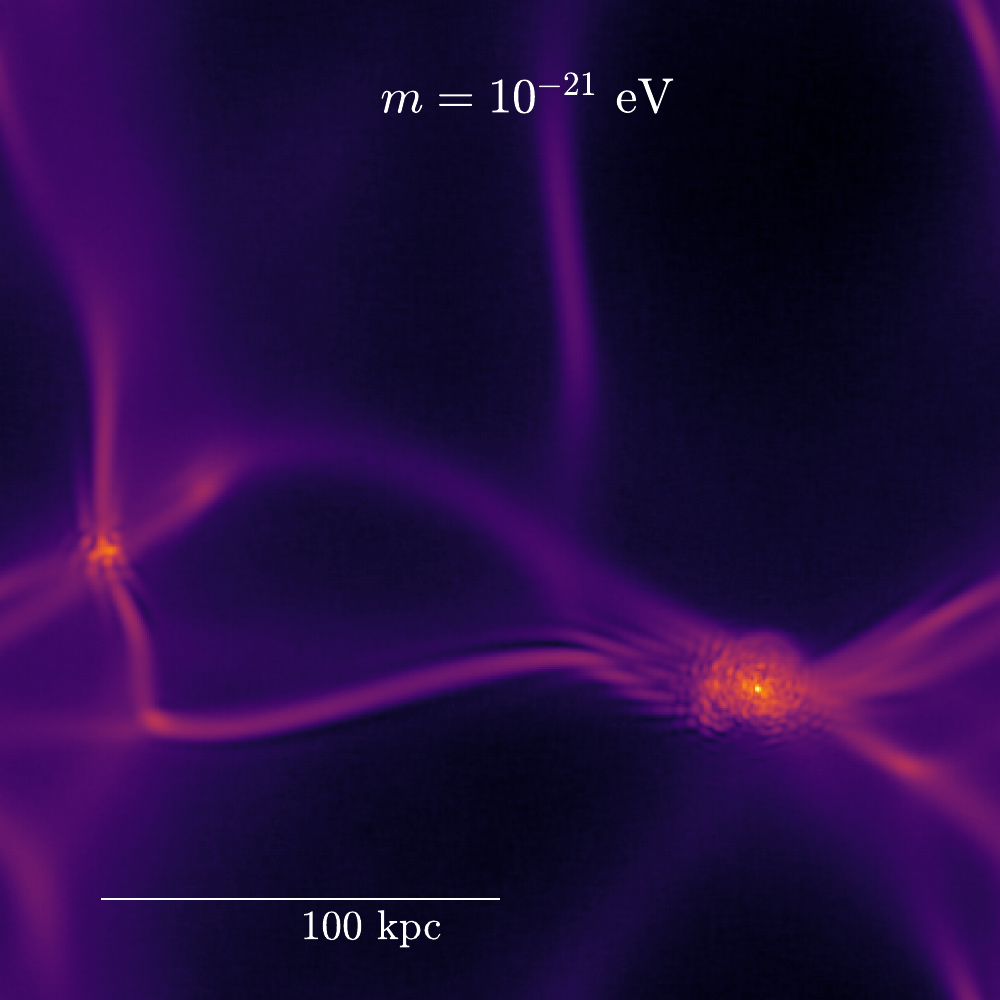}
\includegraphics[width=0.42\textwidth]{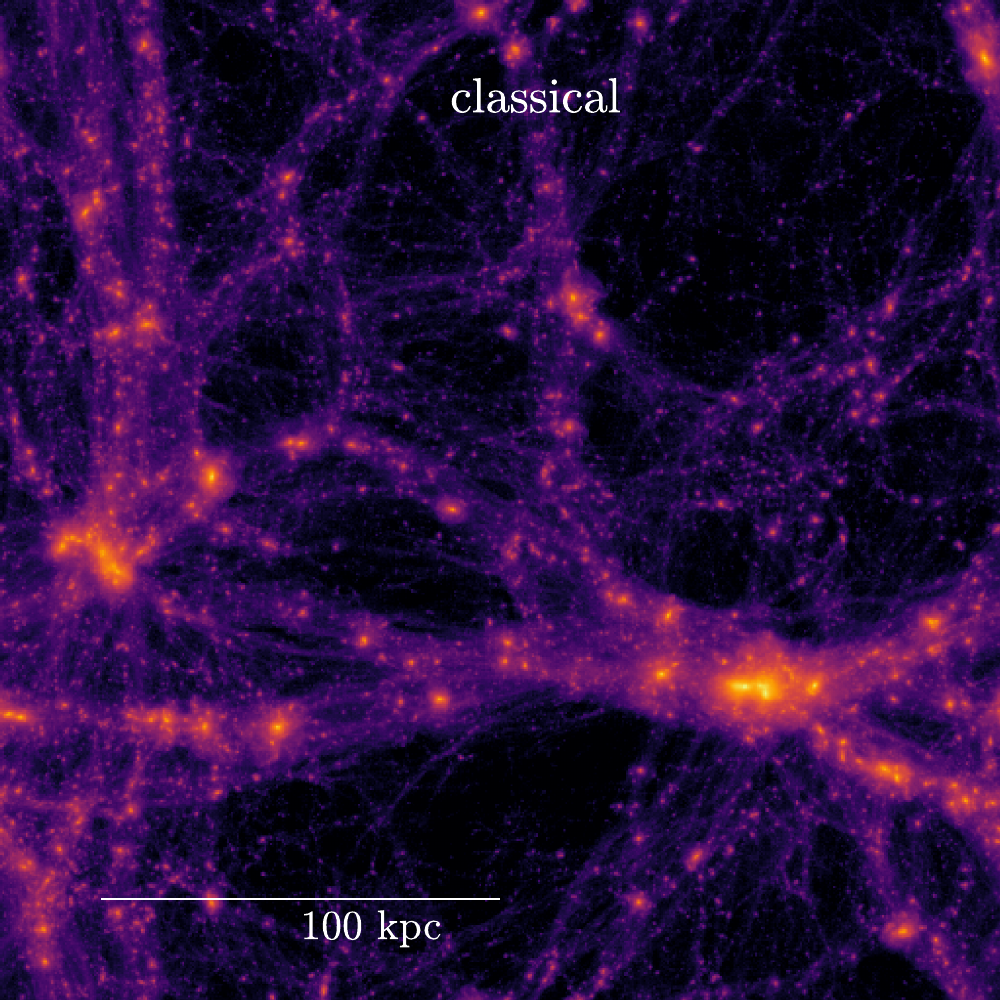} \\
\includegraphics[width=0.3\textwidth]{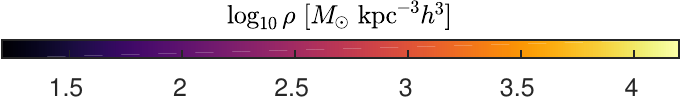} \\
\includegraphics[width=0.5\textwidth]{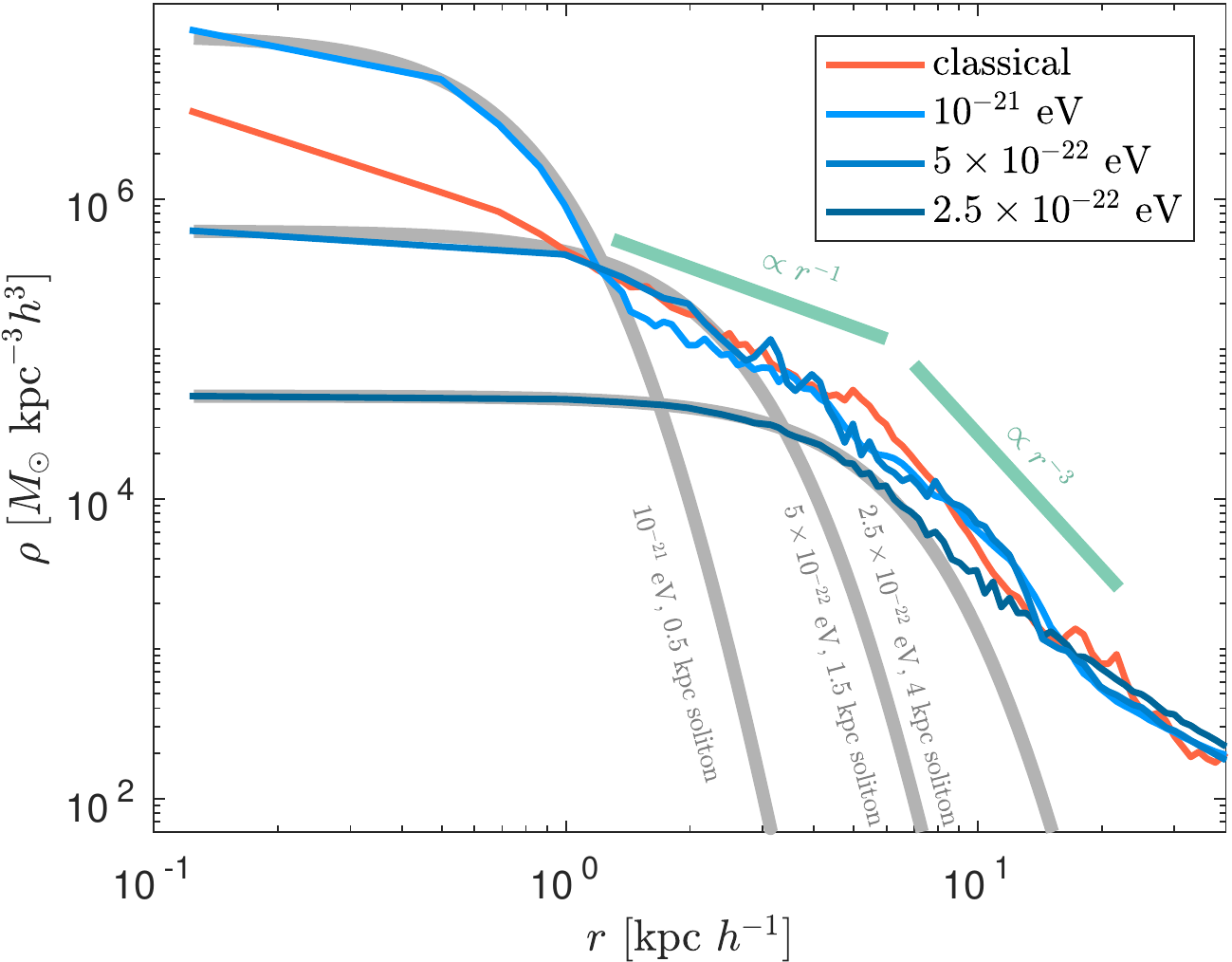}
\end{center}
\caption{A $250h^{-1}~{\rm kpc}$ cosmological simulation at $z=3$, 
evolved either as CDM (VP equations) or as FDM (SP) with axion masses
$m=2.5\cdot 10^{-22}~{\rm eV},5\cdot 10^{-22}~{\rm eV},10^{-21}~{\rm eV}$. Shown are the projected
dark matter density and the radial profile of the most massive halo.
Quantum effects suppress small scale structure (more for small axion masses).
The CDM NFW-like profile of the central halo is approximately recovered for 
larger axion mass, but remains regularized by a central solitonic core.}
\label{fig:a3d}
\end{figure*}

\subsubsection{Setup}

We now consider a modification of 
the previous cold setup by introducing multiple velocities
in the initial conditions.
Rather than setting the initial velocity to $0$, 
the velocity of the particles are distributed 
equally among $13$ phase sheets with velocities:
\begin{equation}
\begin{bmatrix}
(0,0), (1,0),(-1,0),(0,1),(0,-1),  \\
(1,1),(1,-1),(-1,1),(-1,-1), \\
(2,0),(-2,0),(0,2),(0,-2)
\end{bmatrix}
\times v_0
\end{equation}
where $v_0=0.1927~{\rm km}~{\rm s}^{-1}$.

The corresponding wave function is constructed using
Eq.~(\ref{eqn:init}) without the need to add random phase offsets 
because we have a finite number of discrete velocities.
Note that $v_0$ has been chosen so that the individual waves
${\rm e}^{i m \mathbf{x}\cdot\mathbf{v} /\hbar}$ fit periodically into the
domain.

Figure~\ref{fig:b2d} shows the density distribution at $t=1~{\rm Mpc}~({\rm km}~{\rm s}^{-1})^{-1}$.
In this problem, in the classical limit, the Gaussians at the
different velocities move apart
and merge gravitationally in the horizontal direction, leading to thin
horizontal caustic structures.

\subsubsection{Discussion}

The SP equations again capture the classical solution as the boson mass increases, 
with an L1 norm error in the potential that goes as $m^{-2}$ in the asymptotic limit.
This is despite the fact that the density distribution has order unity errors at $t=0$ due to the interference of the multiple phase sheets. 
This is of note because previously only cold initial 
conditions in 2D have been tested \citep{2017arXiv171100140K}.

At low boson mass, the uncertainty principle smoothes out the
rich phase-space structure and the
system collapses into a single soliton like core.
The potential well created is actually significantly deeper than the classical case, 
by conservation of energy, as kinetic energy is lacking inside the collapsed structure. This demonstrates
that the quantum pressure tensor does not necessarily just smooth out structure and leads to less-bound structures, 
so its effect in cosmological simulations needs to be understood using self-consistent simulations.

\subsection{2D warm initial conditions}\label{sec:2dc}

\subsubsection{Setup}

We also consider a modification of our 2D collapse problem 
that uses warm initial conditions.
Namely, the velocities are set to have a Gaussian distribution 
with velocity dispersion $v_{\rm disp}=0.1~{\rm km}~{\rm s}^{-1}$.

Again, the corresponding initial wave function is constructed using
Eq.~(\ref{eqn:init}), 
by sampling the velocities represented on our discretized Fourier grid
(see \cite{1993ApJ...416L..71W} for details of the discrete version of
Eq.~(\ref{eqn:init})).
Here random phases added to each velocity wave component are necessary to prevent 
the different waves from forming coherent peaks/nodes.

We symmetrize the initial conditions across the $x$ and $y$ axis in our 
simulation (both SP and VP case) to keep the center of mass exactly at the
center.

Figure~\ref{fig:c2d}  again shows the density distribution and convergence of the potential at $t=1~{\rm Mpc}~({\rm km}~{\rm s}^{-1})^{-1}$.
In this problem, the two Gaussians merge into a single elongated halo supported against collapse by velocity dispersion.

\subsubsection{Discussion}

Even in the case of warm initial conditions, with random phases added to the different quantum wave modes, 
the solution recovers the classical answer as $m\to\infty$, and the potential 
again converges as $m^{-2}$.
This is good news, as it demonstrates that the Schr\"odinger equations can capture
velocity dispersion in a meaningful way.
The normalization of the L1 norm error is the largest here of the three cases (cold initial conditions show the least error).
In the pre-asymptotic limit (e.g. $m=1.25\times 10^{-22}$~eV)
the merged core does not even form (matter is dispersed and is outside of the zoomed-in region shown in Figure~\ref{fig:c2d}).
As the boson mass is increased, the collapsed structure is puffed up compared to the classical VP limit 
due to the effect of the additional quantum pressure tensor, but in the limit $m\to\infty$ this effect disappears.

The classical solution itself has a flat cored $r^{0}$ profile in 2D, hence no singularity is present and no central 2D soliton core is present in the SP solution.

\subsection{3D cosmological initial conditions}\label{sec:3d}

\subsubsection{Setup}

We perform dark matter only cosmological simulations
of CDM in an expanding universe
under the evolution of VP and SP.
We adopt the following cosmological parameters:
$\Omega_{\rm m}=0.27$, 
$\Omega_{\rm \Lambda}=0.73$, 
$\Omega_{\rm b}=0.046$,
$\sigma_8=0.81$,
$n_{\rm s}=0.96$, 
and $H_0=100~h~{\rm km}~{\rm s}^{-1}~{\rm Mpc}^{-1}$
with $h=0.7$.
Initial conditions for CDM, a realization of a random Gaussian field, are generated at $z=127$ as 
in \cite{2014Natur.509..177V}
and evolved with $N$-body gravitational solver of the {\sc Arepo} code.
We include cosmological expansion in these simulations.
It is interesting to note that the quantum mechanical fluctuations that seed the early universe and, in the CDM paradigm, become classical after inflation remain quantum mechanical here. 
The initial conditions for the wave function are
generated using the strategy outlined in section~\ref{sec:ic}
and using Eq.~(\ref{eqn:ic}) to obtain the phase from the initial single-stream velocity.
The domain is a periodic box of size $250~h^{-1}$~kpc
and the simulation uses a Fourier grid of $1024^3$ cells.
The setup which is designed to accurately resolve fine structure
in the wave function (hence the small box size), does not necessarily capture
large-scale ($>10$~Mpc) structure formation.
Axion dark matter halos are known to contain sub-kiloparsec solitonic cores, 
stable structures at halo centers where the quantum pressure 
tensor supports against gravitational collapse
\citep{2014NatPh..10..496S,2016PhRvD..94d3513S,2017MNRAS.471.4559M}.
Therefore sub-kpc scales need to be resolved. We consider axion
masses $m=2.5\times 10^{-22}~{\rm eV}$, $m=5\times 10^{-22}~{\rm
eV}$, and $m=10^{-21}~{\rm eV}$ for the
simulations.

Figure~\ref{fig:a3d} shows the projected dark matter density distribution at redshift $z=3$. We investigate the radial profile of the most massive halo formed in the simulation, also shown in the figure.

\subsubsection{Discussion}\label{sec:disc3d}

Noticeably, the uncertainty principle erases small-scale
structure in the simulations. In the $m=2.5\times 10^{-22}~{\rm
eV}$ simulation
structure has been reduced to a single spheroid on these scales.
As the axion mass increases, more of the substructure of
scale-free CDM is recovered.

Of particular interest is the halo profiles formed in the simulation.
In CDM, halos are known to have a universal NFW-like profile
 \citep{1996ApJ...462..563N}, which follows an $r^{-1}$
 radial profile in the center and an $r^{-3}$ density profile at the outskirts. Notably, this profile is singular at the center. Note, the NFW profile is simply a numerical fit to simulations, 
 and alternative models that closely resemble it also exist, e.g. \cite{1965TrAlm...5...87E}.

Of interest is whether the SP equations recover the classical NFW-like profile in the limit $m\to\infty$.
The authors of \cite{2014NatPh..10..496S} simulated 
a cosmological box with a small axion mass of $m=2.5\times
10^{-22}~{\rm eV}$, and the halos they found had an $r^{-3}$ outer profile
but the centers were dominated by a soliton core.
The SP equations admit stable soliton solutions, cores supported against collapse under self-gravity by the quantum pressure tensor. 
The soliton core solutions are well-approximated by 
\begin{equation}
\rho_{\rm soliton}(r)\simeq
\rho_0
\left[1+0.091\times \left(\frac{r}{r_{\rm c}}\right)^2\right]^{-8}
\label{eqn:fit}
\end{equation}
\citep{2014PhRvL.113z1302S}, where $r_{\rm c}$ is the core radius in co-moving units, $a$ is the cosmological scale factor at redshift $z=1/a-1$, and $\rho_0$ is the central comoving density given by:
\begin{equation}
\rho_0\simeq 3.1\times 10^{15} a^{-1} 
\left(\frac{2.5\times 10^{-22}~{\rm eV}}{m}\right)^2
\left(\frac{{\rm kpc}}{r_{\rm c}}\right)^4
\frac{M_\odot}{{\rm Mpc}^3}.
\end{equation}
The soliton cores satisfy a mass-radius relationship:
\begin{equation}
M_{\rm c} =  3.59\times 10^7 a^{-1} \left(\frac{2.5\times 10^{-22}~{\rm eV}}{m}\right)^2 \left(\frac{r_{\rm c}}{\rm kpc}\right)^{-1} ~M_\odot.
\label{eqn:mr}
\end{equation}

Figure~\ref{fig:a3d} shows 
that we indeed recover the $r^{-1}$ part of the NFW-like profile
when the axion mass is large. Oscillatory quantum interference patterns have been smoothed out in the radial averaging.
The center of the singularity remains regularized by the soliton core.

Thus, the halo profile starts reasonably resembling the classical collisionless
solution for axion mass of $m=10^{-21}~{\rm eV}$.
We are limited by numerical resolution to explore even larger axion masses,
where we would expect to recover more of the smaller features too, and at the
present axion mass the de Broglie wavelength still smears away many of the
small-scale substructure in the CDM simulation.
Note recently the authors of \cite{2018arXiv180102320L} have simulated fully virialized massive axion dark matter halos (in non-cosmological contexts)
in a regime where the classical limit is not fully recovered and in this case
the profiles beyond the soliton core are found to be best described by a
fermionic King model \citep{2015PhRvD..92l3527C}.
We do see in the radial profile that the profile transition from $r^{-3}$ to $r^{-1}$ is not as sharp in the FDM $m=10^{-21}~{\rm eV}$ case as the classical CDM case. Thus, the profile may be better described as a fermionic King model plus soliton core that approximates the NFW-like classical solution increasingly at larger axion mass.
A discussion is offered in Appendix~\ref{sec_lb}.

We can ask how does the soliton core mass scale with the
axion mass for a fixed halo mass in the limit as the SP equations start approximating the VP equations. 
Importantly, under the SP--VP correspondence, 
the soliton core mass $M_{\rm c}$ should go to $0$
as the axion mass $m\to\infty$, i.e., all quantum phenomena are
expected to disappear as the classical limit is recovered. 
This can be estimated as follows.

In the limit of large axion mass we can predict the
mass of the solitonic core for a given dark matter profile.
Assume the classical limit has a center cuspy profile:
\begin{equation}
\rho(r) = \left(\frac{\rho_{1\rm kpc}}{10^6~M_\odot~{\rm kpc}^{-3}h^3}\right) \left(\frac{r}{{\rm kpc}}\right)^{-1}.
\end{equation}
Then, the 
mass enclosed in a radius $R$ is $M=\int_0^R 4\pi r^2 \rho(r)\,dr \propto R^2 \rho_{1\rm kpc}$. Now replace this region inside radius $R$ with a solitonic core of the same mass. Let $R=3r_{\rm c}$ so that the solitonic core contains $95$~per~cent of its mass inside $R$. The rest of the solution is fixed to the classical limit. 
The soliton core must obey its mass-radius relation, Eq.~(\ref{eqn:mr}).
Hence, the mass of the soliton core needs to be
\begin{multline}
M_{\rm c} = 4.2\times 10^7 
\left(\frac{\rho_{1\rm kpc}}{10^6~M_\odot~{\rm kpc}^{-3}h^3}\right)^{1/3} \\ \times a^{-2/3} 
\left(\frac{2.5\times 10^{-22}~{\rm eV}}{m}\right)^{4/3}~M_\odot
\end{multline}
with corresponding radius
\begin{multline}
r_{\rm c} = 0.85\left(\frac{\rho_{1\rm kpc}}{10^6~M_\odot~{\rm kpc}^{-3}h^3}\right)^{-1/3} \\ \times a^{-1/3}
\left(\frac{2.5\times 10^{-22}~{\rm eV}}{m}\right)^{2/3}~{\rm kpc}
\end{multline}
in the limit that the classical solution starts being recovered (i.e., large enough axion mass, or massive enough halos).
We see that indeed the mass and radius of the regularizing soliton core both shrinks to $0$ as $m\to\infty$.

The above scaling implies that soliton mass to halo mass would scale as
$M_{\rm c}/M_{\rm halo}\propto \Xi^{2/3}$ as the classical limit is approached (large axion mass, large halo mass), 
where 
\begin{equation}
\Xi \equiv \frac{\lvert E_{\rm halo}\rvert/M_{\rm halo}^3}{(Gm/\hbar)^2}
\label{eqn:Xi}
\end{equation}
 is a dimensionless parameter characterizing a halo \cite{2017MNRAS.471.4559M}, 
$E_{\rm halo}\sim GM_{\rm halo}^2/R_{\rm halo} \propto M_{\rm halo}^{5/3}$ the
halo total energy. This implies $M_c\propto M_{\rm
halo}^{1/9}$. $\Xi$ is invariant under the scaling symmetries of the SP
equation. For a halo dominated by quantum effects, (e.g. primarily composed of
the soliton core), $\Xi\sim 1$ whereas a halo resembling the classical limit has
$\Xi \ll 1$. Thus, more precisely, 
the $2/3$ scaling of soliton core mass to halo mass may be expected to hold for $\Xi \ll 1$.
Note, in contrast, for fully virialized halos over the range $\Xi\in [10^{-4},10^{-2}]$, \cite{2017MNRAS.471.4559M} found 
$M_{\rm c}/M_{\rm halo}\propto \Xi^{1/3}$ (implying
$M_c\propto M_{\rm
halo}^{5/9}$). These
halos were not close to the classical limit.
And in a cosmological context (not fully virialized),
Ref. \cite{2014NatPh..10..496S} 
found $M_{\rm c}/M_{\rm halo}\propto \Xi^{1/2}$ (implying
$M_c\propto M_{\rm
halo}^{1/3}$).
Again, the halos considered here were only over a limited mass range $M_{\rm halo}\sim 10^9$--$10^{11}~M_\odot$ with small axion mass
$m=2.5\times 10^{-22}~{\rm eV}$, thus again these halos did not
reach $\Xi \ll 1$.

Interestingly, the soliton mass may be comparable to the masses of supermassive black holes at the centers of halos for small axion masses, which may have important astrophysical consequences or observable signatures.

\section{Concluding Remarks}
\label{sec:conc}

Our numerical experiments demonstrate that a correspondence
between the SP and VP equations exists in the sense
that for the wide range of complex test problems we simulated (caustics, multi phase-sheet, warm conditions, cosmological simulations), the potential converges to the classical solution as $(\hbar/m)^2$.
Hence the force field is also converged to the classical answer in the limit $\hbar/m\to 0$, despite the fact that
the density field, riddled with order unity quantum interference patterns, does
not converge to the classical limit. The convergence of the time-evolved
potential allows the emergence of classical behavior as $\hbar\to 0$. Our
simulations support the emergence of classical behavior in the $\hbar\to 0$
limit up an (perhaps unobservable) oscillation in the density, which may be due
to the fact that one is working with one-particle wave-function or condensates.
A more general quantum state could possibly prevent this feature and could also 
avoid the coarse graining through the Husimi transformation
\citep{2017PhRvD..96h3504F}.

As a consequence, in the case of axion dark matter, since the force-field converges, the evolution of baryons coupled to the dark matter through the potential only thus will recover the CDM solution as the axion mass $m\to\infty$. Indeed, our simulations show that we can start to recover NFW-like profiles in axion dark matter cosmological simulations for large axion masses
$m\gtrsim 10^{-21}$~eV. Thus standard QCD 
axions ($m\sim 10^{-6}$--$10^{-3}$~eV) may be a very
natural candidate for CDM.

The SP equations, despite having a single fluid velocity at a given point, can capture multiple phase-sheets shell-crossings  and warm velocity distributions since the Schr\"odinger equations
are linear wave equations that can superimpose different wave speeds, and the amplitude of the resulting interference oscillations are suppressed in the potential that governs the non-linear time evolution.

Furthermore, the SP equations regularize singularities (caustics) of the VP equations.
For example, the singularity in the NFW halo profile of CDM is replaced by a soliton core (a non-linear quantum structure).

Given a classical 6D distribution function, we have demonstrated one may construct from it a 3D wave function that preserves its information. 
However, the reverse may not always be possible. 
For example, the wave function is allowed to have a discontinuity in its phase, 
which corresponds to a classical velocity of $\infty$.
Such initial conditions lead to non-classical behavior: for example, 
two self-gravitating cores with opposite phases can bounce off each other rather than directly merge under self-gravity due to the large quantum pressure between them \citep{2016PhRvD..94d3513S}.
Thus in this sense, the SP equations encode a richer set of behaviors than the VP equations, encompassing both the classical and quantum limit.

Of recent interest has been the construction of optical analogs in the laboratory with physics governed by the SP equations \citep{2016NatCo...713492R}. 
The mentioned study explored the dynamics of rotating boson stars using a continuous-wave laser to pump a slab of lead-doped glass.
One may envision the design of optical analogs of cosmological systems (CDM and FDM) as well. Through the SP--VP correspondence, one may probe behavior of both the classical and quantum regime.

A formal mathematical proof of the full range of conditions under which the SP--VP correspondence hold remains to be carried out, but numerically we have shown important properties of its nature in the time evolution of complicated systems relevant to cosmology.

\section*{Acknowledgments}
Support for this work was provided by NASA through Einstein Postdoctoral Fellowship grant number PF7-180164 awarded by the \textit{Chandra} X-ray Center, which is operated by the Smithsonian Astrophysical Observatory for NASA under contract NAS8-03060.
The authors acknowledge the Texas Advanced Computing Center (TACC) at The University of Texas at Austin for providing HPC resources that have contributed to the research results reported within this paper. URL: \url{http://www.tacc.utexas.edu}.
Some of the computations in this paper were run on the Odyssey cluster supported by the FAS Division of Science, Research Computing Group at Harvard University.
The authors would like to thank Michael Kopp, David Spergel, Jerry Ostriker, Jim Stone, Aaron Szasz, Sauro Succi, Scott Tremaine, and Pavel  Friedrich for discussions. 
\bibliography{mybib}{}

\begin{thebibliography}{10}

\bibitem{2016NatCo...713492R}
T.~{Roger} {\em et~al.},
\newblock Nature Communications {\bf 7}, 13492 (2016).

\bibitem{harrison2016quantum}
P.~Harrison and A.~Valavanis,
\newblock {\em Quantum wells, wires and dots: theoretical and computational
  physics of semiconductor nanostructures} (John Wiley \& Sons, 2016).

\bibitem{1968PhRv..172.1331K}
D.~J. {Kaup},
\newblock Physical Review {\bf 172}, 1331 (1968).

\bibitem{1969PhRv..187.1767R}
R.~{Ruffini} and S.~{Bonazzola},
\newblock Physical Review {\bf 187}, 1767 (1969).

\bibitem{2003CQGra..20R.301S}
F.~E. {Schunck} and E.~W. {Mielke},
\newblock Classical and Quantum Gravity {\bf 20}, R301 (2003).

\bibitem{2017PhRvD..96b4002S}
N.~{Sennett}, T.~{Hinderer}, J.~{Steinhoff}, A.~{Buonanno}, and S.~{Ossokine},
\newblock \prd {\bf 96}, 024002 (2017), 1704.08651.

\bibitem{1989PhRvA..39.4207M}
M.~{Membrado}, A.~F. {Pacheco}, and J.~{Sa{\~n}udo},
\newblock \pra {\bf 39}, 4207 (1989).

\bibitem{1994PhRvD..50.3650S}
S.-J. {Sin},
\newblock \prd {\bf 50}, 3650 (1994), hep-ph/9205208.

\bibitem{1994PhRvD..50.3655J}
S.~U. {Ji} and S.~J. {Sin},
\newblock \prd {\bf 50}, 3655 (1994), hep-ph/9409267.

\bibitem{1996PhRvD..53.2236L}
J.-W. {Lee} and I.-G. {Koh},
\newblock \prd {\bf 53}, 2236 (1996), hep-ph/9507385.

\bibitem{2000PhRvL..85.1158H}
W.~{Hu}, R.~{Barkana}, and A.~{Gruzinov},
\newblock Physical Review Letters {\bf 85}, 1158 (2000), astro-ph/0003365.

\bibitem{2000ApJ...534L.127P}
P.~J.~E. {Peebles},
\newblock \apjl {\bf 534}, L127 (2000), astro-ph/0002495.

\bibitem{2000NewA....5..103G}
J.~{Goodman},
\newblock \na {\bf 5}, 103 (2000), astro-ph/0003018.

\bibitem{2002esas.book..165M}
T.~{Matos}, F.~S. {Guzm{\'a}n}, L.~A. {Ure{\~n}a-L{\'o}pez}, and
  D.~{N{\'u}{\~n}ez},
\newblock {\em {Scalar Field Dark Matter}} (Kluwer Academic Publishers, 2002).

\bibitem{2007JCAP...06..025B}
C.~G. {B{\"o}hmer} and T.~{Harko},
\newblock \jcap {\bf 6}, 025 (2007), 0705.4158.

\bibitem{2011PhRvD..84d3531C}
P.-H. {Chavanis},
\newblock \prd {\bf 84}, 043531 (2011), 1103.2050.

\bibitem{2014PhRvL.113z1302S}
H.-Y. {Schive} {\em et~al.},
\newblock Physical Review Letters {\bf 113}, 261302 (2014), 1407.7762.

\bibitem{2014NatPh..10..496S}
H.-Y. {Schive}, T.~{Chiueh}, and T.~{Broadhurst},
\newblock Nature Physics {\bf 10}, 496 (2014), 1406.6586.

\bibitem{2016PhRvD..94d3513S}
B.~{Schwabe}, J.~C. {Niemeyer}, and J.~F. {Engels},
\newblock \prd {\bf 94}, 043513 (2016), 1606.05151.

\bibitem{2017MNRAS.471.4559M}
P.~{Mocz} {\em et~al.},
\newblock \mnras {\bf 471}, 4559 (2017), 1705.05845.

\bibitem{2017PhRvD..95d3541H}
L.~{Hui}, J.~P. {Ostriker}, S.~{Tremaine}, and E.~{Witten},
\newblock \prd {\bf 95}, 043541 (2017), 1610.08297.

\bibitem{2017arXiv170405057L}
J.-W. {Lee},
\newblock ArXiv e-prints  (2017), 1704.05057.

\bibitem{moore1994}
B.~{Moore},
\newblock \nat {\bf 370}, 629 (1994).

\bibitem{flores1994}
R.~A. {Flores} and J.~R. {Primack},
\newblock \apjl {\bf 427}, L1 (1994), arXiv:astro-ph/9402004.

\bibitem{2010AdAst2010E...5D}
W.~J.~G. {de Blok},
\newblock Advances in Astronomy {\bf 2010}, 789293 (2010), 0910.3538.

\bibitem{boylan-kolchin2011}
M.~{Boylan-Kolchin}, J.~S. {Bullock}, and M.~{Kaplinghat},
\newblock \mnras {\bf 415}, L40 (2011), 1103.0007.

\bibitem{2012MNRAS.422.1203B}
M.~{Boylan-Kolchin}, J.~S. {Bullock}, and M.~{Kaplinghat},
\newblock \mnras {\bf 422}, 1203 (2012), 1111.2048.

\bibitem{2015PhRvD..91j3512H}
R.~{Hlozek}, D.~{Grin}, D.~J.~E. {Marsh}, and P.~G. {Ferreira},
\newblock \prd {\bf 91}, 103512 (2015), 1410.2896.

\bibitem{1977PhRvL..38.1440P}
R.~D. {Peccei} and H.~R. {Quinn},
\newblock Physical Review Letters {\bf 38}, 1440 (1977).

\bibitem{1927ZPhy...40..322M}
E.~{Madelung},
\newblock Zeitschrift fur Physik {\bf 40}, 322 (1927).

\bibitem{1993ApJ...416L..71W}
L.~M. {Widrow} and N.~{Kaiser},
\newblock \apjl {\bf 416}, L71 (1993).

\bibitem{2017arXiv171100140K}
M.~{Kopp}, K.~{Vattis}, and C.~{Skordis},
\newblock ArXiv e-prints  (2017), 1711.00140.

\bibitem{2017arXiv171004846G}
M.~{Garny} and T.~{Konstandin},
\newblock ArXiv e-prints  (2017), 1710.04846.

\bibitem{2008gady.book.....B}
J.~{Binney} and S.~{Tremaine},
\newblock {\em {Galactic Dynamics: Second Edition}} (Princeton University
  Press, 2008).

\bibitem{1996ApJ...471..385C}
P.~H. {Chavanis}, J.~{Sommeria}, and R.~{Robert},
\newblock \apj {\bf 471}, 385 (1996).

\bibitem{1967MNRAS.136..101L}
D.~{Lynden-Bell},
\newblock \mnras {\bf 136}, 101 (1967).

\bibitem{husimi1940}
K.~{Husimi},
\newblock Proc. Phys. Math. Soc. Japan {\bf 22}, 264 (1940).

\bibitem{1989PhRvA..40.2894S}
R.~T. {Skodje}, H.~W. {Rohrs}, and J.~{Vanbuskirk},
\newblock \pra {\bf 40}, 2894 (1989).

\bibitem{2015PhRvE..91e3304M}
P.~{Mocz} and S.~{Succi},
\newblock \pre {\bf 91}, 053304 (2015), 1503.03869.

\bibitem{2017MNRAS.465.3154M}
P.~{Mocz} and S.~{Succi},
\newblock \mnras {\bf 465}, 3154 (2017), 1611.02757.

\bibitem{2010MNRAS.401..791S}
V.~{Springel},
\newblock \mnras {\bf 401}, 791 (2010), 0901.4107.

\bibitem{2003AmJPh..71..483B}
F.~A. {Barone}, H.~{Boschi-Filho}, and C.~{Farina},
\newblock American Journal of Physics {\bf 71}, 483 (2003), quant-ph/0205085.

\bibitem{2016iqm..book.....G}
D.~J. {Griffiths},
\newblock {\em {Introduction to Quantum Mechanics}} (Pearson Prentice Hall,
  2016).

\bibitem{2014Natur.509..177V}
M.~{Vogelsberger} {\em et~al.},
\newblock \nat {\bf 509}, 177 (2014), 1405.1418.

\bibitem{1996ApJ...462..563N}
J.~F. {Navarro}, C.~S. {Frenk}, and S.~D.~M. {White},
\newblock \apj {\bf 462}, 563 (1996), astro-ph/9508025.

\bibitem{1965TrAlm...5...87E}
J.~{Einasto},
\newblock Trudy Astrofizicheskogo Instituta Alma-Ata {\bf 5}, 87 (1965).

\bibitem{2018arXiv180102320L}
S.-C. {Lin}, H.-Y. {Schive}, S.-K. {Wong}, and T.~{Chiueh},
\newblock ArXiv e-prints  (2018), 1801.02320.

\bibitem{2015PhRvD..92l3527C}
P.-H. {Chavanis}, M.~{Lemou}, and F.~{M{\'e}hats},
\newblock \prd {\bf 92}, 123527 (2015), 1409.7840.

\bibitem{2017PhRvD..96h3504F}
P.~{Friedrich} and T.~{Prokopec},
\newblock \prd {\bf 96}, 083504 (2017).

\bibitem{1986MNRAS.219..285T}
S.~{Tremaine}, M.~{Henon}, and D.~{Lynden-Bell},
\newblock \mnras {\bf 219}, 285 (1986).

\bibitem{1998MNRAS.300..981C}
P.-H. {Chavanis},
\newblock \mnras {\bf 300}, 981 (1998).

\end{thebibliography}

\appendix

\section{Violent relaxation of collisionless self-gravitating
systems}
\label{sec_lb}

Here we discuss important concepts of violent relaxation of collisionless
self-gravitating systems, relevant to understand what is
happening in the
cosmological simulations of dark matter halos. We also address
a preliminary generalization of these concepts to the case of bosonic
particles.

\subsection{Classical systems}

The VP equations have a very complicated dynamics associated with 
phase mixing and nonlinear Landau damping. As a result, they develop
intermingled filaments at smaller and smaller scales and a coarse-grained
description becomes necessary to smooth out this intricate filamentation. 
Starting from an out-of-equilibrium initial condition, a collisionless
self-gravitating system  generically
experiences a process of violent relaxation. While the
fine-grained distribution
function $f({\bf x},{\bf v},t)$ always evolves in time,
the coarse-grained distribution
function reaches a
quasistationary state $\overline{f}_{\rm QSS}({\bf x},{\bf v})$
on a very short
timescale of the order of the 
dynamical time $t_D$. This is because, for $t\gtrsim t_D$,
the evolution of  $f({\bf x},{\bf v},t)$ takes place at a scale smaller
than the coarse-graining mesh. Using arguments of
statistical mechanics, Lynden-Bell \cite{1967MNRAS.136..101L} predicted that the
coarse-grained 
distribution function in the QSS should be of the form
\footnote{This expression
is valid when the fine-grained distribution function takes only two values
$f=\eta_0$ and $f=0$. In the general case, one has to discretize the initial
condition in several levels $\eta$. The equilibrium coarse-grained 
distribution function is then a sum of distributions of the form of
Eq. (\ref{lb1}) on the
different levels \cite{1967MNRAS.136..101L}.} 
\begin{eqnarray}
\label{lb1}
\overline{f}_{\rm LB}({\bf x},{\bf
v})=\frac{\eta_0}{1+e^{\beta(\epsilon-\mu)}},
\end{eqnarray}
where $\epsilon=v^2/2+V({\bf x})$ is the individual energy
of the
particles, $\mu$ is the chemical potential, $\eta_0$ is the maximum value of
the fine-grained distribution
function
$f({\bf x},{\bf v},t)$  and $\beta=\eta_0/T_{\rm eff}$
is an effective inverse temperature. Note that $T_{\rm eff}$ has not the
dimension of a
temperature but $T_{\rm eff}/\eta_0$ has the dimension of a velocity
dispersion.
We note that the mass $m$ of the particles does not appear in
Lynden-Bell's theory which is based on the Vlasov equation since the system is
collisionless.

The Lynden-Bell distribution (\ref{lb1}) is the most probable
state, or
most mixed state, taking into account all the constraints of the Vlasov
equation. It is similar to the Fermi-Dirac distribution. A sort of exclusion
principle similar to the Pauli exclusion principle
in
quantum mechanics (but with another interpretation) arises in the theory of
Lynden-Bell due to the incompressibility of the flow in phase space and the
conservation of the distribution function (on the fine-grained scale) by the
Vlasov equation. As a result, the
coarse-grained distribution function must
always be smaller than the maximum value of the fine-grained distribution
function ($\overline{f}\le \eta_0$). This is why the Lynden-Bell
statistics  is similar to the Fermi-Dirac
statistics.\footnote{Actually, the  Lynden-Bell statistics corresponds to a
fourth type of statistics where the particles experience an exclusion principle
but are distinguishable \cite{1967MNRAS.136..101L}. However, this again leads to
a distribution
function similar to the Fermi-Dirac distribution.} This suggests that the
process of
violent relaxation is similar in some respect to the relaxation of
self-gravitating fermionic
particles.

In the nondegenerate limit ($\overline{f}_{\rm LB}\ll\eta_0$), the  Lynden-Bell
distribution function reduces to 
\begin{eqnarray}
\label{lb2}
\overline{f}_{\rm LB}({\bf x},{\bf v})=\eta_0 e^{-\beta
(\epsilon-\mu)},
\end{eqnarray}
which is similar to the Maxwell-Boltzmann statistics. The Lynden-Bell theory
of violent relaxation explains how a collisionless self-gravitating system
``thermalizes'' on a very short timescale (much shorter than the two-body
relaxation time $\sim Nt_D$).\footnote{It is shell crossing that
permits violent relaxation or dynamical phase mixing to take place. In
cosmology, after the development of the linear Jeans instability, overdensity
regions collapse until the collapse is reversed by a collective particle
``bounce'' at pericenter. Such counterstreaming leads to the well-known
collisionless two-stream instabilty followed by Landau damping. After a few
oscillations, the configuration settles into a stable equilibrium
state. The fluctuations of the gravitational potential lead
to particle thermalization.  Violent relaxation serves to
``thermalize'' the initial velocity profile. Shell crossing and phase mixing
generate a velocity dispersion where there was none to begin with.}

The Lynden-Bell distribution (\ref{lb1}) function suffers from a defect. At
large distances
the system is always nondegenerate implying that the density decreases as
$\rho\propto r^{-2}$ as for a classical self-gravitating isothermal gas. As a
result,  the total mass is infinite. The physical solution to this ``infinite
mass problem'' is to invoke incomplete relaxation \cite{1967MNRAS.136..101L}. In
practice, the
system may
not mix sufficiently well and the QSS reached by the system may differ from the
Lynden-Bell prediction \cite{1986MNRAS.219..285T}. Another limitation of the
Lynden-Bell
distribution
function is that it does not take  into account the
escape of high energy particles. This problem can be cured by developing a
dynamical description of the process of violent relaxation.

A question of fundamental interest is to derive the dynamical  equation that
governs the evolution of the coarse-grained distribution function
$\overline{f}({\bf x},{\bf v},t)$. Writing 
$f=\overline{f}+\delta f$ and $V=\overline{V}+\delta V$
and taking the
local average of the Vlasov equation we get
\begin{eqnarray}
\label{lb3}
\frac{\partial \overline{f}}{\partial t}+{\bf v}\cdot
\frac{\partial
\overline{f}}{\partial {\bf
x}}-\nabla\overline{V}\cdot \frac{\partial \overline{f}}{\partial {\bf
v}}=\frac{\partial}{\partial {\bf v}}\cdot \overline{\delta f\nabla\delta V}.
\end{eqnarray}
This equation shows that the correlations of the fluctuations of the
gravitational potential and distribution function create an effective
``collision'' term. Using heuristic arguments
based on a Maximum Entropy Production Principle (MEPP), the authors of
\cite{1996ApJ...471..385C} have
proposed a relaxation
equation of the form
\begin{eqnarray}
\label{lb4}
&&\frac{\partial \overline{f}}{\partial t}+{\bf v}\cdot
\frac{\partial
\overline{f}}{\partial {\bf
x}}-\nabla\overline{V}\cdot \frac{\partial \overline{f}}{\partial {\bf
v}}\nonumber\\
&=&\frac{\partial}{\partial {\bf v}}\left\lbrack D\left
(\frac{\partial
\overline{f}}{\partial {\bf
v}}+\beta \overline{f}(1-\overline{f}/\eta_0){\bf v}\right )\right\rbrack
\end{eqnarray}
which takes into account the Lynden-Bell exclusion principle ($\overline{f}\le
\eta_0$). More elaborated
expressions of the effective collision term derived from kinetic
theory are given in \cite{1996ApJ...471..385C,1998MNRAS.300..981C}. Equation
(\ref{lb4}) is similar to a
fermionic Fokker-Planck
equation. From this equation, one can derive a truncated Lynden-Bell
distribution \cite{1998MNRAS.300..981C}:
\begin{eqnarray}
\label{lb5}
\overline{f}=A\frac{e^{-\beta(\epsilon-\epsilon_m)}-1}{1+\frac{A}{\eta_0}e^{
-\beta(\epsilon-\epsilon_m)}}\qquad (\epsilon\le \epsilon_m), 
\end{eqnarray}
\begin{eqnarray}
\label{lb6}
\overline{f}=0\qquad (\epsilon\ge \epsilon_m). 
\end{eqnarray}
There is a truncation above a certain escape energy $\epsilon_m$ which takes
into account tidal effects when $\epsilon_m<0$ or simply the escape of unbound
particles with
positive energy  when $\epsilon_m=0$. Eq. (\ref{lb5}) is
called the  fermionic
King model. In the
nondegenerate limit, Eq. (\ref{lb5}) is similar to the classical King model
which was
introduced in relation to globular clusters evolving under the effect of
two-body encounters.  However, in the present context, the thermalization of the
system
is due to Lynden-Bell's type of
relaxation and the fermionic nature of the distribution function is related to
Lynden-Bell's exclusion principle. The fermionic King model (\ref{lb5}) has a
finite mass.
It has been studied in detail in \cite{2015PhRvD..92l3527C}. It usually displays
a core-halo
structure
with
a degenerate core similar to a ``fermion ball'' (a polytrope of index $n=3/2$)
and a pseudo isothermal halo truncated at the tidal radius. 

Taking the hydrodynamic moments of the coarse-grained Vlasov equation
(\ref{lb4}), we
obtain a system of equations similar to the Jeans
equations but including dissipative effects:
\begin{equation}
\label{lb7}
\frac{\partial\rho}{\partial t}+\nabla\cdot (\rho {\bf u})=0,
\end{equation}
\begin{equation}
\label{lb8}
\frac{\partial {\bf u}}{\partial t}+({\bf u}\cdot \nabla){\bf
u}=-\frac{1}{\rho}\partial_jP_{ij}-\nabla V-\int D\beta
\overline{f}(1-\overline{f}/\eta_0){\bf v}\, d{\bf v}.
\end{equation}
 They are called the damped
Jeans equations. The effective collision term in Eq. (\ref{lb4}) provides 
a source of relaxation
which allows one to
compute the pressure tensor and the friction term  in the Jeans equations by
using a local thermodynamical
equilibrium assumption:
\begin{eqnarray}
\label{lb9}
\overline{f}_{\rm LTE}({\bf x},{\bf
v},t)=\frac{\eta_0}{1+e^{\beta
({\bf v}-{\bf u}({\bf x},t))^2/2+\alpha({\bf x},t)}}.
\end{eqnarray}
This is a manner to close the hierarchy of moment equations. This leads to 
a system of hydrodynamic equations of the form
\begin{equation}
\label{lb10}
\frac{\partial\rho}{\partial t}+\nabla\cdot (\rho {\bf u})=0,
\end{equation}
\begin{equation}
\label{lb11}
\frac{\partial {\bf u}}{\partial t}+({\bf u}\cdot \nabla){\bf
u}=-\frac{1}{\rho}\nabla P_{\rm LB}-\nabla V-\xi{\bf u},
\end{equation}
called the damped Euler equations.\footnote{For
simplicity, we have neglected degeneracy effects in the friction force (see
\cite{1996ApJ...471..385C} for generalization) and introduced the friction
coefficient
$\xi=D\beta$ satisfying an Einstein-like relation.} In these equations, $P_{\rm
LB}(\rho)$ is the
equation of state associated with the
Lynden-Bell distribution (\ref{lb1}).\footnote{More generally $P(\rho)$ should
be computed from the fermionic King distribution.
Alternatively, it can be simply approximated by $P=\rho T_{\rm
eff}/\eta_0+(1/5)[3/(4\pi\eta_0)]^{2/3}\rho^{5/3}$ where
the $n=3/2$ polytropic equation of state 
$P=(1/5)[3/(4\pi\eta_0)]^{2/3}\rho^{5/3}$  describes the pseudo-fermionic core
and the
linear equation of state $P=\rho T_{\rm
eff}/\eta_0$ describes the isothermal halo. } It coincides with
the
Fermi-Dirac
equation of state where $gm^4/h^3$ ($g$ is the
multiplicity of the quantum states) is replaced by $\eta_0$.
This pressure law
includes an effective temperature term and also takes into account the
Lynden-Bell exclusion
principle. On the other hand, the friction term $-\xi {\bf u}$ may be related to
a form of nonlinear Landau
damping. More elaborated hydrodynamic equations are given in
\cite{1996ApJ...471..385C}.

\begin{figure}
\begin{center}
\includegraphics[width=0.45\textwidth]{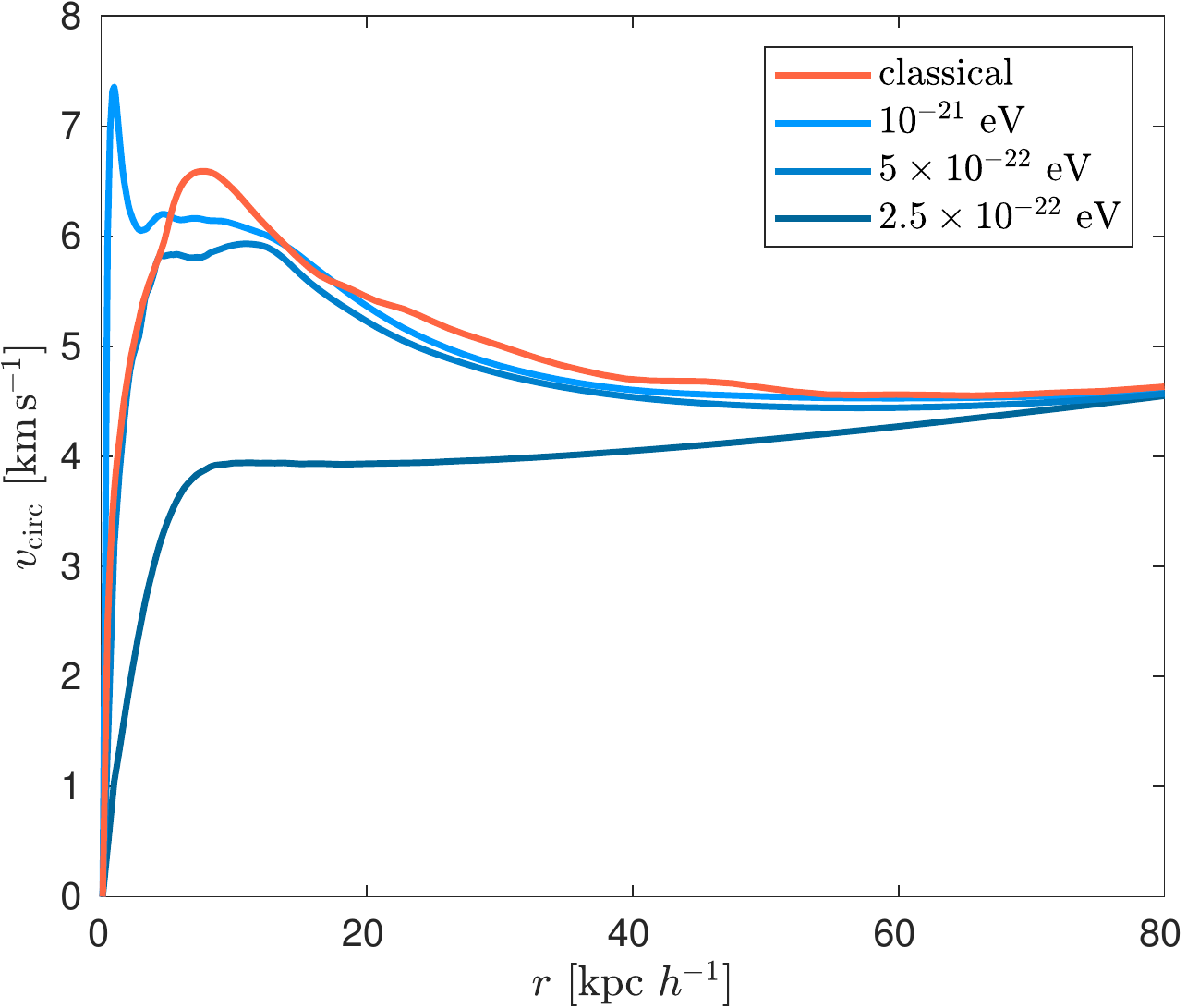}
\end{center}
\caption{Rotation curves for the halos of Fig.~\ref{fig:a3d} simulated with different axion masses. Flat rotation curves are achieved, and large axion masses. For small axion masses, the effect of the solitonic core is clearly visible.
}
\label{fig:rotcurve}
\end{figure}

\begin{figure}
\begin{center} 
\includegraphics[width=0.45\textwidth]{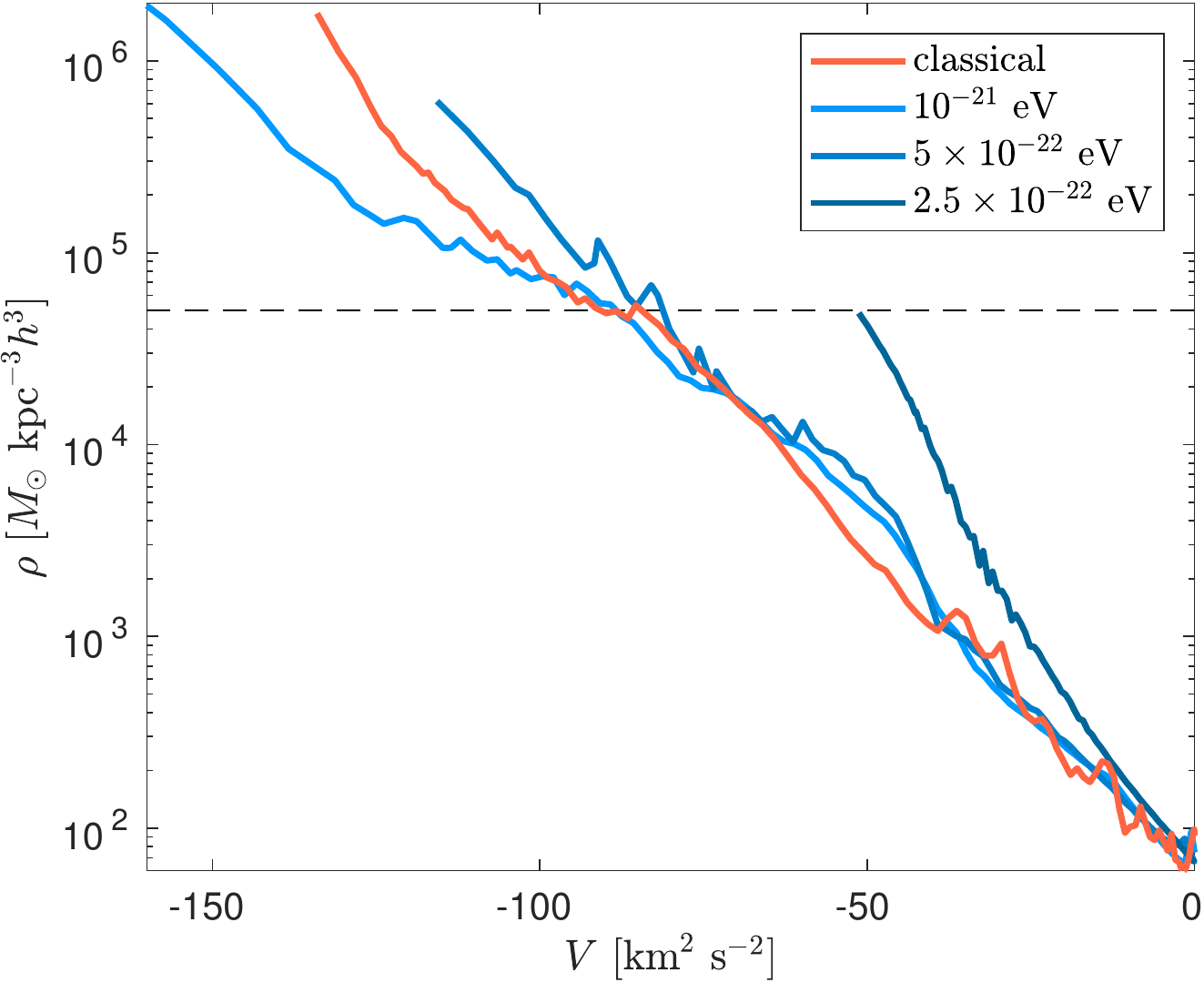}
\end{center}
\caption{Log density versus potential for the halos of Fig.~\ref{fig:a3d} simulated with different axion masses. A linear relation is indicative of a pseudo-isothermal halo. At potential centers the FDM simulations differ from the classical solution due to the presence of the soliton core.}
\label{fig:potential}
\end{figure}

\subsection{Quantum systems made of bosons}

Considering now self-gravitating FDM (i.e., dark matter evolved
with the SP equations), we argue that the Wigner-Poisson 
equations
should also lead to a process of violent relaxation. This process is
probably very complex to describe theoretically since, already at the classical
level, it
is not fully understood. Combining the previous ideas, we can heuristically
consider a general hydrodynamic
model of the form
\begin{equation}
\label{lb12}
\frac{\partial\rho}{\partial t}+\nabla\cdot (\rho {\bf u})=0,
\end{equation}
\begin{equation}
\label{lb13}
\frac{\partial {\bf u}}{\partial t}+({\bf u}\cdot \nabla){\bf
u}=-\nabla
V_Q-\frac{1}{\rho}\nabla P_{\rm
LB}-\nabla V-\xi{\bf u},
\end{equation}
which combines the properties of Eqs. (\ref{eqn:madQ}) and (\ref{lb11}). The
steady states
of these equations lead to
a system presenting a core-halo structure with a solitonic core and an
isothermal
halo. The halo, being collisionless, is independent of the
boson mass $m$ while the solitonic core, having a quantum origin, strongly
depends on $m$ and disappears when $\hbar/m\rightarrow 0$. There should also be
a degenerate pseudo fermionic core, similar to a
fermion ball (polytrope of index $n=3/2$), due to Lynden-Bell's
exclusion principle.  This fermion ball
disappears in the nondegenerate limit.

\subsection{Discussion}
\label{sec_discussion}

Considering the classical CDM model, we note that the QSS obtained in numerical
simulations is given by an NFW-like profile which
is incompatible with the Lynden-Bell
distribution. Indeed, in the nondegenerate limit, the
statistical theory of Lynden-Bell predicts  an isothermal profile which has a
flat core and a $\rho\propto r^{-2}$ halo, while the numerical NFW profile
displays a $\rho\propto r^{-1}$ central cusp and a  $\rho\propto r^{-3}$ halo.
The reason why Lynden-Bell's theory does not work in that context is not clearly
understood.

Considering now the FDM model, it seems that
quantum mechanics  favors the establishment of the Lynden-Bell distribution by
eliminating the $r^{-1}$ cusp and replacing it by a solitonic core. Our
simulations in Sec. \ref{sec:3d} show that the halo (away from the soliton) is
not very different from an isothermal halo. Indeed, at large distances, the
density
profile is not inconsistent with the $\rho\propto r^{-2}$ isothermal 
law (it actually works better than the $\rho\propto r^{-3}$
fit).\footnote{This may be due to small box size in the
simulations. In larger simulations, the profile decays as  $r^{-3}$ implying
that this steepening may be an effect of tidal interactions.
This is in agreement
with the King model which is a pseudo isothermal distribution modified by tidal
effects \cite{2015PhRvD..92l3527C}.} This leads to an almost flat circular
velocity profile at large distances (see Fig.~\ref{fig:rotcurve}). Even more
convincingly, if we plot
$\ln\rho$ vs $V$ (see Fig.~\ref{fig:potential}) we observe a reasonable linear
relationship in the halo which is consistent with the Boltzmann law $\rho\propto
e^{-\beta V}$ obtained from Eq. (\ref{lb2}). Finally, we note that
\cite{2017arXiv171100140K} recently found that the
QSS,
in addition to containing a solitonic core (of quantum mechanical origin), is
close to the Lynden-Bell
distribution, or to the fermionic
King model \cite{1998MNRAS.300..981C,2015PhRvD..92l3527C}. Therefore, the 
halo of FDM appears to be close to isothermal.
This apparent isothermal halo may be justified by the process of
 collisionless violent relaxation.

Thus a FDM halo behaves as follows.
The halo is characterized by the parameter $\Xi$ of Eq.~(\ref{eqn:Xi}) discussed
in section~\ref{sec:disc3d}, which 
defines how much quantum mechanics affects the halo structure
(this parameter is generally larger for small axion masses or low halo masses).
$\Xi \sim 1$ is a halo that is a pure solitonic core (the pure quantum
mechanical limit). If the axion mass is increased in the
cosmological simulation, 
then $\Xi$ decreases, bringing
the halos more towards the classical limit.
For $\Xi \lesssim 1$ the halo starts resembling the fermionic King plus soliton
core model.
As $\Xi \ll 1$, this profile approaches the classical
solution obtained in a CDM simulation.

\vfill\eject
\end{document}